# Статистическое исследование стенограмм подсчета голосов на многомандатных выборах: Выявление электоральных фальсификаций и реконструкция итогов голосования

**А.В. Подлазов[1], В. Макаров[2,3]**

*[1] Институт прикладной математики им. М.В. Келдыша РАН*
*[2] Виговский исследовательский центр квантовой связи, университет г. Виго;*
*[3] член территориальной избирательной комиссии пос. Власиха Московской области с правом решающего голоса*

**Аннотация.** Предложены новые методы электоральной статистики. С их помощью исследованы стенограммы подсчета голосов по выборам муниципальных депутатов. Построены и апробированы эффективные статистические тесты, позволяющие обнаруживать вброс пачек фальшивых бюллетеней как на уровне отдельных кандидатов, так и бюллетеней как целого. Изучено различие между фальсификациями за власть и против оппозиции. Сконструированы индикаторы, оценивающие правдоподобие стенограмм в целом. Предложена модель поведения избирателей, позволяющая реконструировать итоги многомандатного голосования на уровне отдельного участка. Выполнена независимая проверка реконструкции.

# Statistical study of the transcript of vote counts in multi-member constituencies: Identifying electoral fraud and reconstructing voting returns

**A.V. Podlazov[1], V. Makarov[2,3]**

*[1] RAS Keldysh Institute of Applied Mathematics*
*[2] Vigo Quantum Communication Center, University of Vigo;*
*[3] voting member of Territorial election commission of Vlasikha town, Moscow region*

**Abstract.** We propose new methods of electoral statistics. With their help, we study transcripts of vote counting in municipal elections. We construct and apply effective statistical tests to detect the ballot stuffing at the level of individual candidates and ballots as a whole. We study the difference between falsifications for the administration and against the opposition. We construct indicators assessing the credibility of transcripts as a whole. We propose a model of voter behavior to reconstruct the results of multi-member voting at the level of an individual precinct. We carry out an independent verification of the reconstruction.

## Введение

При изучении итогов голосования в случае подозрения на электоральные фальсификации ключевую роль играет формальный подход [1,2,3]. Он делает анализ выборов общедоступным, избавляя от необходимости обращаться к уполномоченным инстанциям, которые давали бы оценку свидетельским показаниям и вещественным доказательствам.

В основе формального подхода лежит статистическое исследование больших массивов данных, порождаемых выборными процедурами. Традиционно такими данными были официальные итоги голосования с обширного набора избирательных участков, публикуемые избирательными комиссиями. Настоящая работа распространяет область приложения электоральной статистики на стенограммы подсчета голосов с отдельных участков.

Одним из наиболее распространенных способов фальсификации итогов голосования является вброс пачек бюллетеней, заполненных не избирателями, а фальсификаторами. Если бюллетени из таких пачек перед подсчетом голосов не перемешивать с обычными бюллетенями, то в стенограмме возникают аномальные области, где фальшивые бюллетени идут подряд. И даже если перемешать бюллетени, но не очень тщательно, то остаются целыми отдельные фрагменты вброшенной пачки бюллетеней, что тоже поддается обнаружению.

Возможности вброса принципиально расширило введение летом 2020 г. трехдневного голосования. Подсчет голосов выполняется в его последний день, а бюллетени предыдущих дней до этого момента должны храниться в запечатанных сейф-пакетах либо опечатанных на ночь урнах. Однако фальсификаторы научились как добавлять бюллетени в урны по ночам, так и незаметно вскрывать сейф-пакеты и подменять их содержимое [4].

### Предмет исследования

Здесь рассматриваются итоги голосования по выборам IV созыва Совета депутатов городского округа Власиха Московской области, прошедших 06-08 сентября 2024 г. Население Власихи составляет 28,6 тыс. чел., из которых избирателей, имеющих право голоса на муниципальных выборах – 17,2 тыс.

Всего в местный Совет входят 15 депутатов – по 5 от 3 избирательных округов. В каждом округе за мандаты боролись по 11 кандидатов (совпадение случайно) на трех обычных (номерных) избирательных участках и одном общем участке *дистанционного электронного голосования* (ДЭГ). И если для последнего доступны сведения только по динамике явки и общим итогам, то для обычных участков имеются полные стенограммы подсчета голосов, подготовленные по его видеозаписям, которые вели наблюдатели на участках. Исключением является участок 213, для которого велась лишь звукозапись, начатая с опозданием и не включающая оглашение первых 17 бюллетеней.

Рассматриваемые выборы – многомандатные, т.е. действительны бюллетени, в которых отмечены от 1 до 5 кандидатов. Такой характер выборов радикально увеличивает объем информации, содержащейся в стенограммах.

Особенности голосования на разных участках

На рис. 1, основанном на данных ресурса https://losevpeter.ru, представлено накопленным итогом по округам число избирателей, проголосовавших дистанционно. На основном графике приведены числа для последовательных четвертьчасовых интервалов, а на врезке – для скользящих часовых интервалов с пятиминутным шагом. Отчетливо видны выраженные пики активности электронных избирателей утром 1-го дня голосования (пятница – единственный рабочий день из трех) и в обеденный перерыв этого дня, свидетельствующие об административном принуждении к голосованию. Это означает, что электронные избиратели более лояльны по сравнению с обычными.

Общую картину рис. 1 обобщают и дополняют первые две строки табл. 1. Величины в них для номерных участков рассчитаны как отношение числа бюллетеней, выданных избирателям на участке в соответствующий день голосования [**4**] (из актов, составляемых при запечатывании сейф-пакетов), и общего числа бюллетеней, выданных избирателям за все три дня (из официальных итогов голосования с сайта http://www.moscow-reg.izbirkom.ru/izbirkom_pages). Как можно видеть, бумажное голосование по сравнению с электронным намного более равномерно во времени. Избыточное голосование в 1-й день связано с нагоном административно-зависимого электората, большой долей которого известны участки 214 и (в особенности) 218, оказавшиеся здесь в лидерах.

Следующие три строки табл. 1 опираются на видеозаписи официальных процедур, выполнявшихся на избирательных участках [4]. Вопросительные знаки означают неполноту записей (213) или их недостаточное качество (214, 219 и 220), что не позволяет установить, вскрывались ли пакеты и перемешивались ли бюллетени перед подсчетом голосов (в этом случае знак «+» или «–» в таблице является результатом дальнейшего анализа). Восклицательные знаки означают, что фальсификаторы действовали открыто, помещая пакеты в сейф незапечатанными (218) или особо тщательно перемешивая бюллетени (217). Общая картина такова: на 8 из 9 номерных избирательных участков сейф-пакеты вскрывались, что компрометирует их содержимое, а умышленное пере-

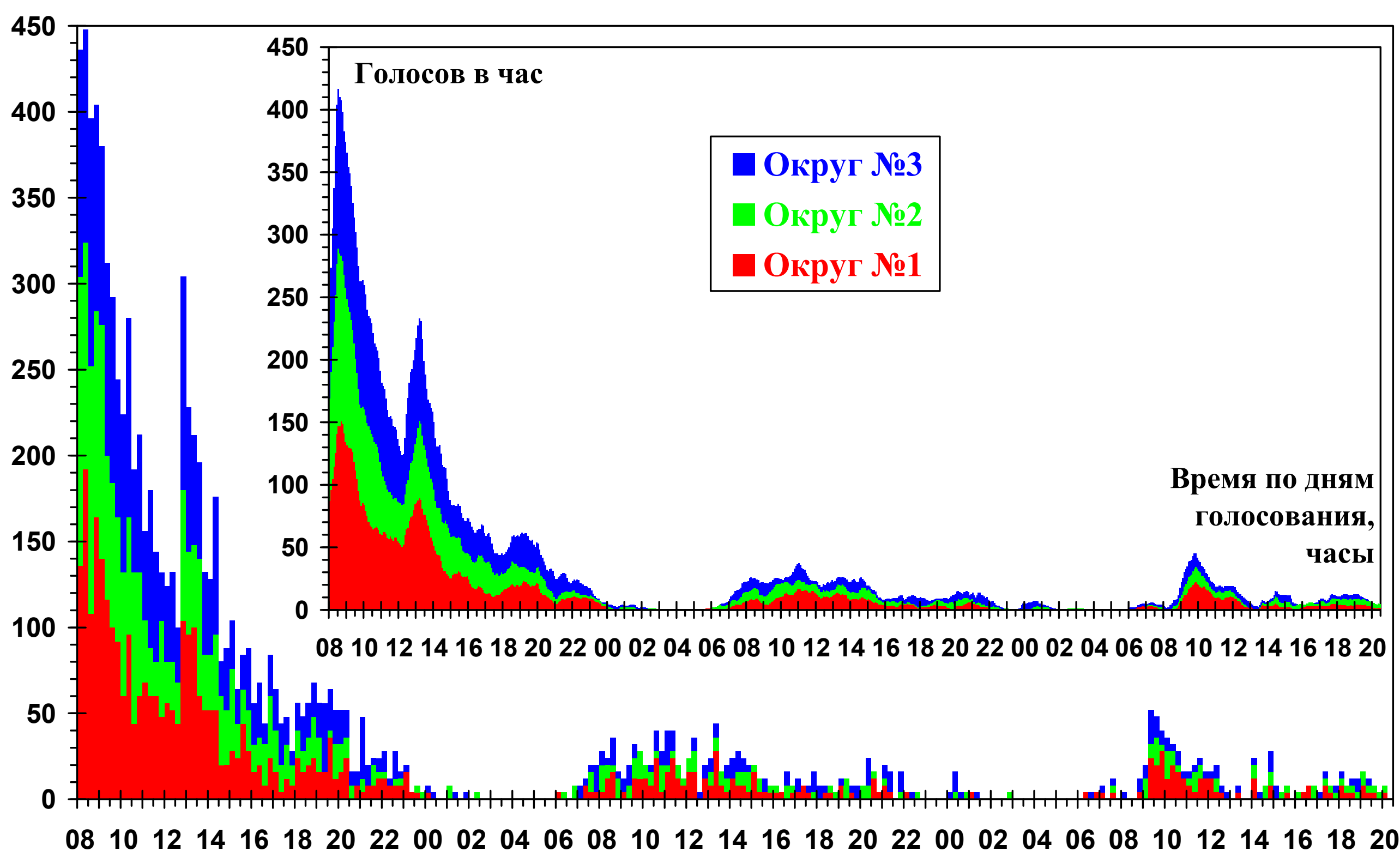

Рис. 1. Динамика дистанционного голосования по часам для всех трех дней мешивание было достаточно частым явлением, чтобы не сомневаться в том, что многие фальсификаторы осознавали необходимость маскировать последствия своих действий, разбивая пачки одинаковых бюллетеней, идущих подряд.

В предпоследней строке табл. 1 приведена доля отметок в бюллетенях, которые проставлены за кандидатов, одержавших победу в округе, измеряемая от общего числа отметок (по официальным итогам). Для участков ДЭГ эта доля ниже, чем для обычных, за исключением участка 215, что подтверждает выводы как о фальсификациях на остальных номерных участках, так и о большей лояльности электронных избирателей по сравнению с обычными.

Участки существенно разняться по масштабу фальсификаций. На участке 218 они были тотальны, что вполне корреспондирует и с высокой лояльностью тамошних избирателей, и с неприкрытыми действиями фальсификаторов. Схожая картина наблюдается и на участке 214, который тоже отличается повышенной лояльностью избирателей, и на участке 217, где фальсификаторы почти преуспели в маскировке своих действий посредством перемешивания бюллетеней. А вот на участках 212, 213, 216, 219 и 220 фальсификации были более-менее умеренными. Можно предположить, что здесь фальсификаторам не хватило либо запаса подменных бюллетеней, либо возможности спокойно работать (в частности, на участок 219, который расположен в прямой видимости от участка 220, во 2-й день голосования вызывалась полиция, что могло напугать фальсификаторов).

Наконец, в последней строке табл. 1 приведено среднее число отметок, сделанных в действительном бюллетене (по официальным итогам). Для элек-тронных участков оно держится на сравнительно низком уровне 3,12÷3,22, для честного участка 215 поднимается до 4,06, а для всех нечестных участков, кроме одного, оказывается еще выше – в диапазоне 4,14÷4,86. Из этой схемы выбивается лишь участок 212, где бюллетень в среднем содержит 3,88 отметок. Это означает, что фальсификации на данном участке осуществлялись не так, как на остальных, что требует отдельного анализа, который выполнен далее.

Таблица 1. Особенности голосования и подведения его итогов по участкам

| Избирательный округ | №1 | | | | №2 | | | | №3 | | | |
|---|---|---|---|---|---|---|---|---|---|---|---|---|
| Избирательный участок | 212 | 213 | 214 | ДЭГ | 215 | 216 | 217 | ДЭГ | 218 | 219 | 220 | ДЭГ |
| Голосование в 1-й день | 35% | 36% | 45% | 79% | 37% | 29% | 37% | 82% | 73% | 36% | 36% | 83% |
| Голосование во 2-й день | 26% | 24% | 26% | 12% | 27% | 27% | 23% | 12% | 18% | 25% | 24% | 11% |
| Вскрытие пакетов 1-го дня | + | +? | +? | — | – | + | + | — | +! | + | +? | — |
| Вскрытие пакетов 2-го дня | + | + | + | — | – | + | + | — | +! | –? | –? | — |
| Перемешивание | + | +? | – | — | – | + | +! | — | – | – | – | — |
| Доля отметок за победителей | 64% | 67% | 84% | 55% | 43% | 69% | 79% | 48% | 96% | 56% | 64% | 51% |
| Отметок на бюллетень | 3,88 | 4,14 | 4,52 | 3,12 | 4,06 | 4,46 | 4,62 | 3,22 | 4,86 | 4,35 | 4,21 | 3,14 |

### Стенограммы подсчета голосов

На рис. 2-10 визуализированы стенограммы подсчета голосов для участков 212-220 [4]. Кандидаты на всех рисунках отсортированы по убыванию их официального результата на соответствующем участке. Черные прямоугольнички означают наличие в бюллетене отметки в поддержку кандидата, а белые – ее отсутствие. Бюллетени изображены в порядке их оглашения при подсчете голосов, причем учитываются только действительные бюллетени.

Также на рисунках приведены синтетические стенограммы для консолидированного содержимого бюллетеней. При *консолидации бюллетеня* весь набор отметок в нем трактуется как единая отметка, способ определения которой зависит от предполагаемого образа действий фальсификаторов. Здесь рассматриваются всего два его варианта – простой и сложный.

Если мы полагаем, что фальсификации имели целью продвижение провластных кандидатов, то весь бюллетень как целое считается поданным «за» при условии, что в нем имеются 5 отметок за кандидатов, набравших наибольшее количество голосов на участке. Это *простые фальсификации*, по-видимому, имевшие место на участках 213-214 и 216-220 (рис. 3-4 и 6-10). Для сравнения этот же способ консолидации применен и к участку 215 (рис. 5).

Если мы полагаем, что фальсификации имели целью сдерживание оппозиционных кандидатов, то весь бюллетень как целое считается поданным «за» при условии, что в нем нет отметок ни за кого из 5 кандидатов, набравших наименьшее количество голосов на участке. Это *сложные фальсификации*, по-видимому, имевшие место только на участке 212 (рис. 2). Именно сложный характер фальсификаций обуславливает низкое среднее число отметок для бюллетеней с этого участка, т.к. вброшенные бюллетени в этом случае могут иметь менее 5 отметок.

На рисунках для облегчения восприятия построены графики локального различия последовательных бюллетеней стенограммы. Этих графики показывают среднее расстояние по Хэммингу (число кандидатов, для которых не совпадают отметки в бюллетенях) $\langle\rho\rangle$ между всеми парами бюллетенями в скользящем окне ширины 25. Чем больше в нем идентично или схоже заполненных бюллетеней, тем ниже среднее расстояние между ними.

Для 11 баллотирующихся кандидатов и от 1 до 5 разрешенных отметок в бюллетене всего возможны 1 023 различных варианта его заполнения. Если вариант $i$ встретился в $n_i$ бюллетенях из их общего числа $n$, то удельная (в пересчете на бюллетень) информационная энтропия по Шеннону составляет $I = -\sum_{i=1}^{1\,023} f_i \log_2 f_i$, где $f_i = n_i / n$. Ее значения, также указанные на рисунках, позволяют судить о том, насколько разнообразно заполнены бюллетени в целом. В частности, если избиратели равновероятно выбирают один из двух списков кандидатов, голосуя за всех его членов, то $I = 1$, а если избиратели действуют абсолютно случайно, то $I \approx 10$. Однако более точные ориентиры на основе априорной информации получить затруднительно, в силу чего энтропию следует рассматривать лишь как относительную, а не абсолютную характеристику участка. С другой стороны, перемешивание бюллетеней никак не влияет на энтропию, благодаря чему ее величина позволяет судить именно о фальсификациях безотносительно попыток их замаскировать.

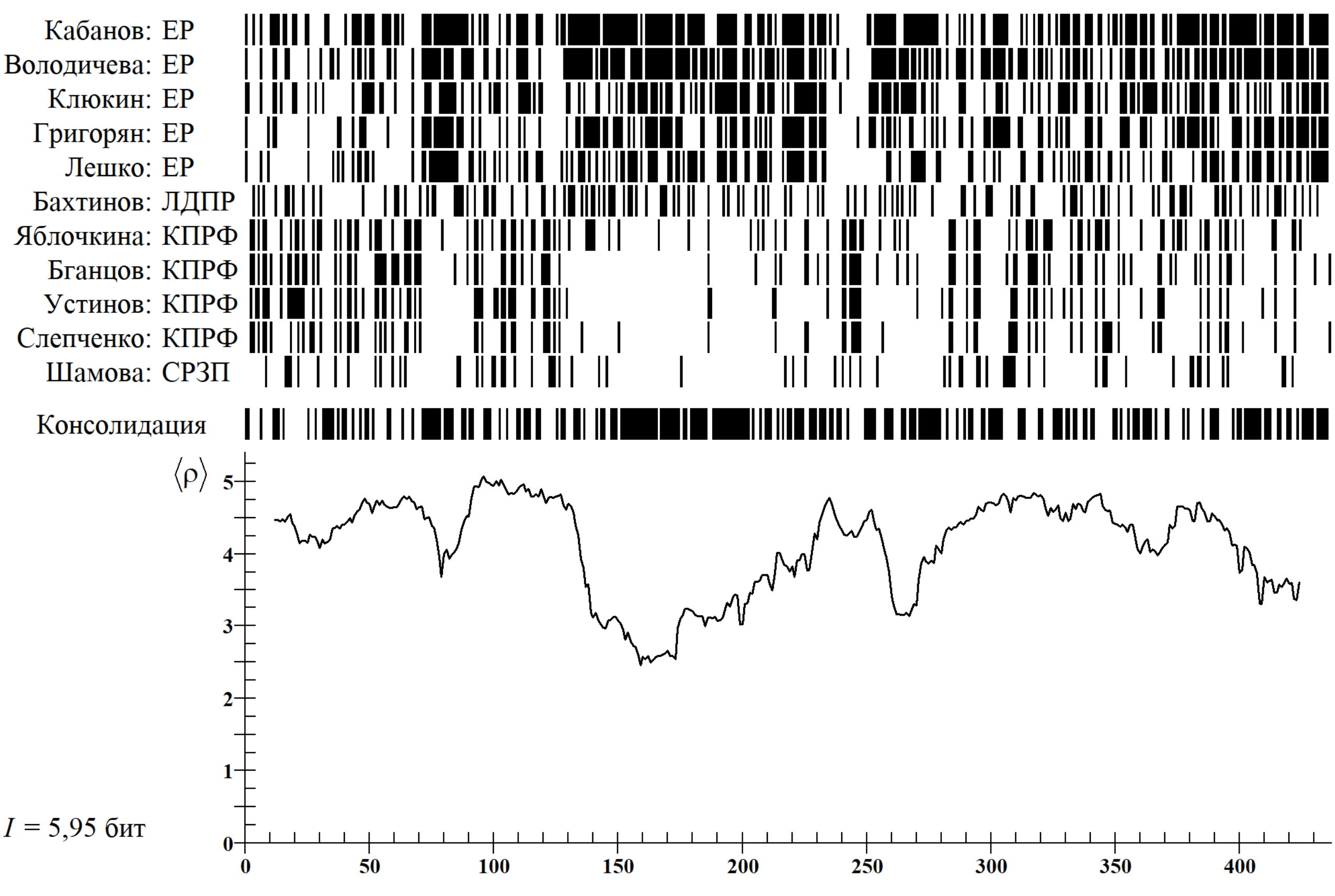

Рис. 2. Стенограмма подсчета голосов для участка 212

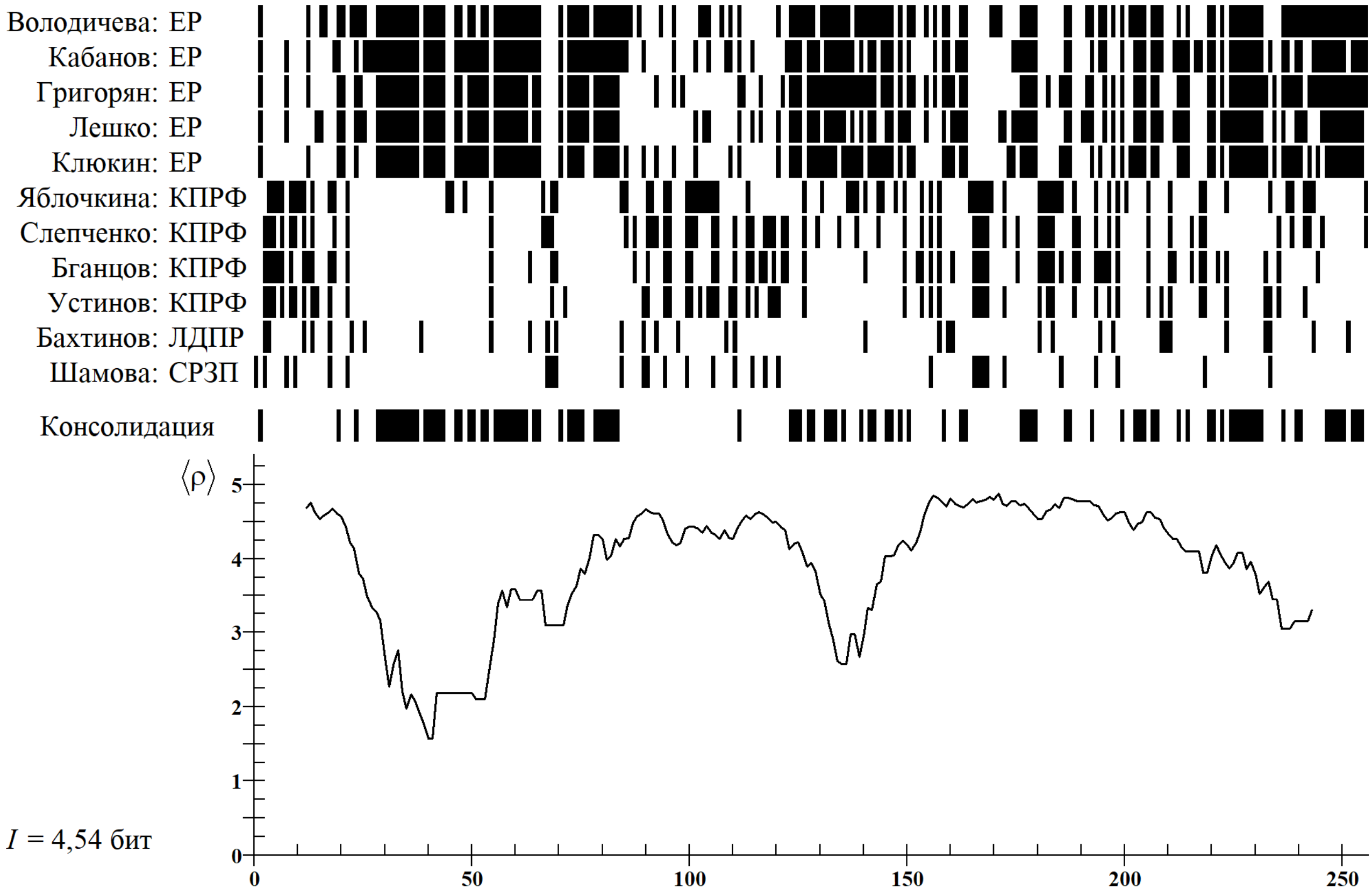

Рис. 3. Стенограмма подсчета голосов для участка 213

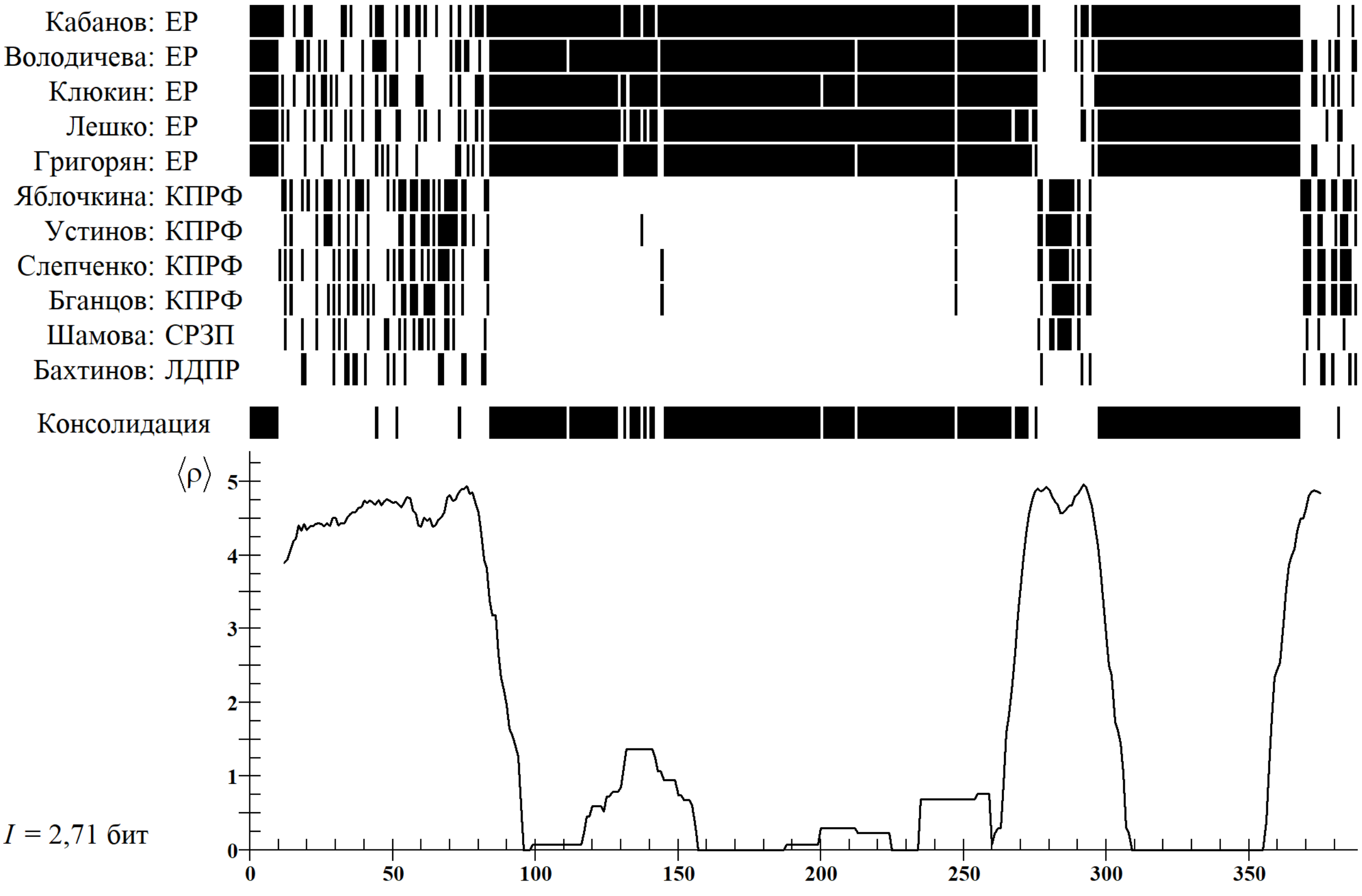

Рис. 4. Стенограмма подсчета голосов для участка 214

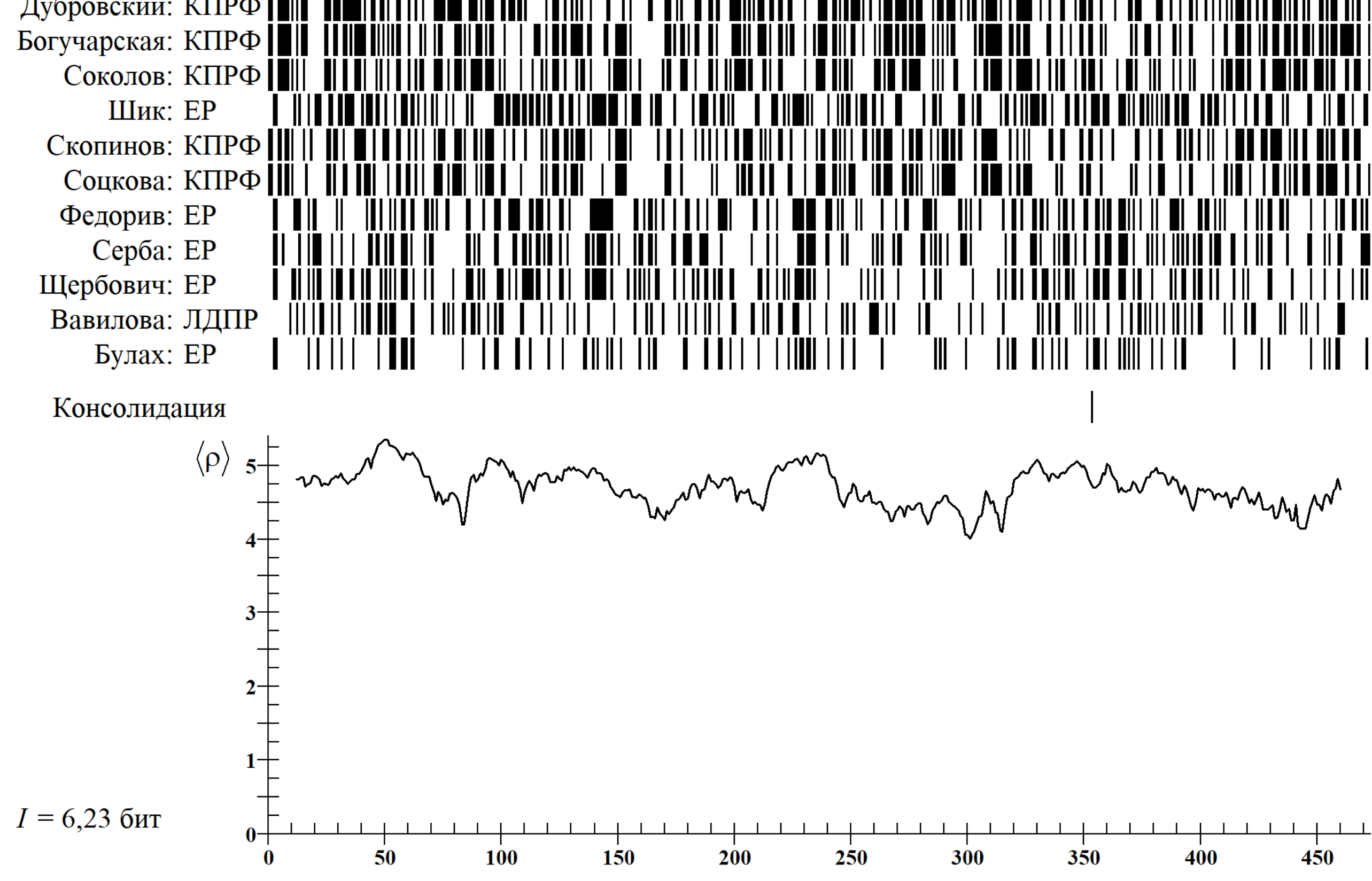

Рис. 5. Стенограмма подсчета голосов для участка 215

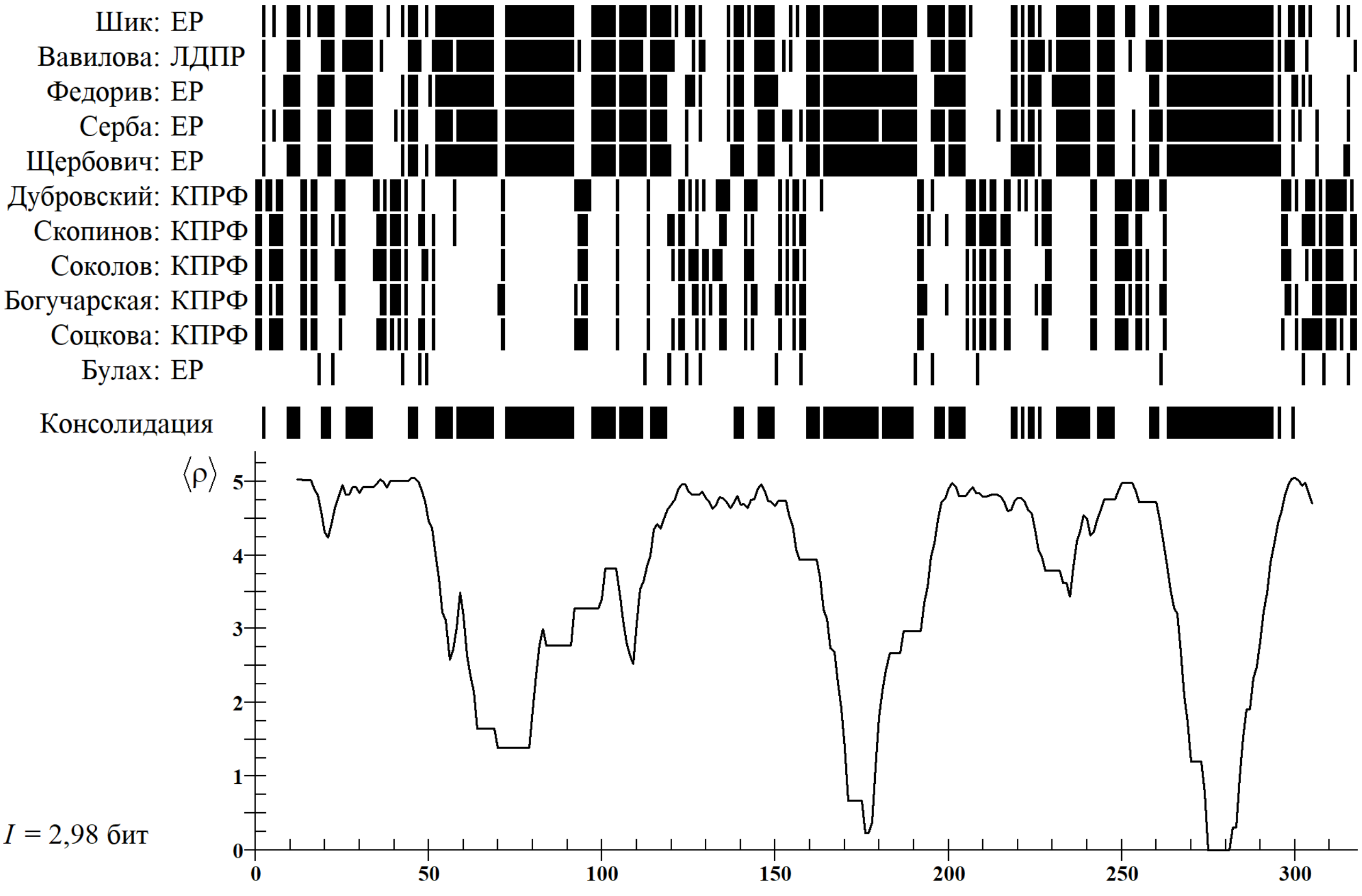

Рис. 6. Стенограмма подсчета голосов для участка 216

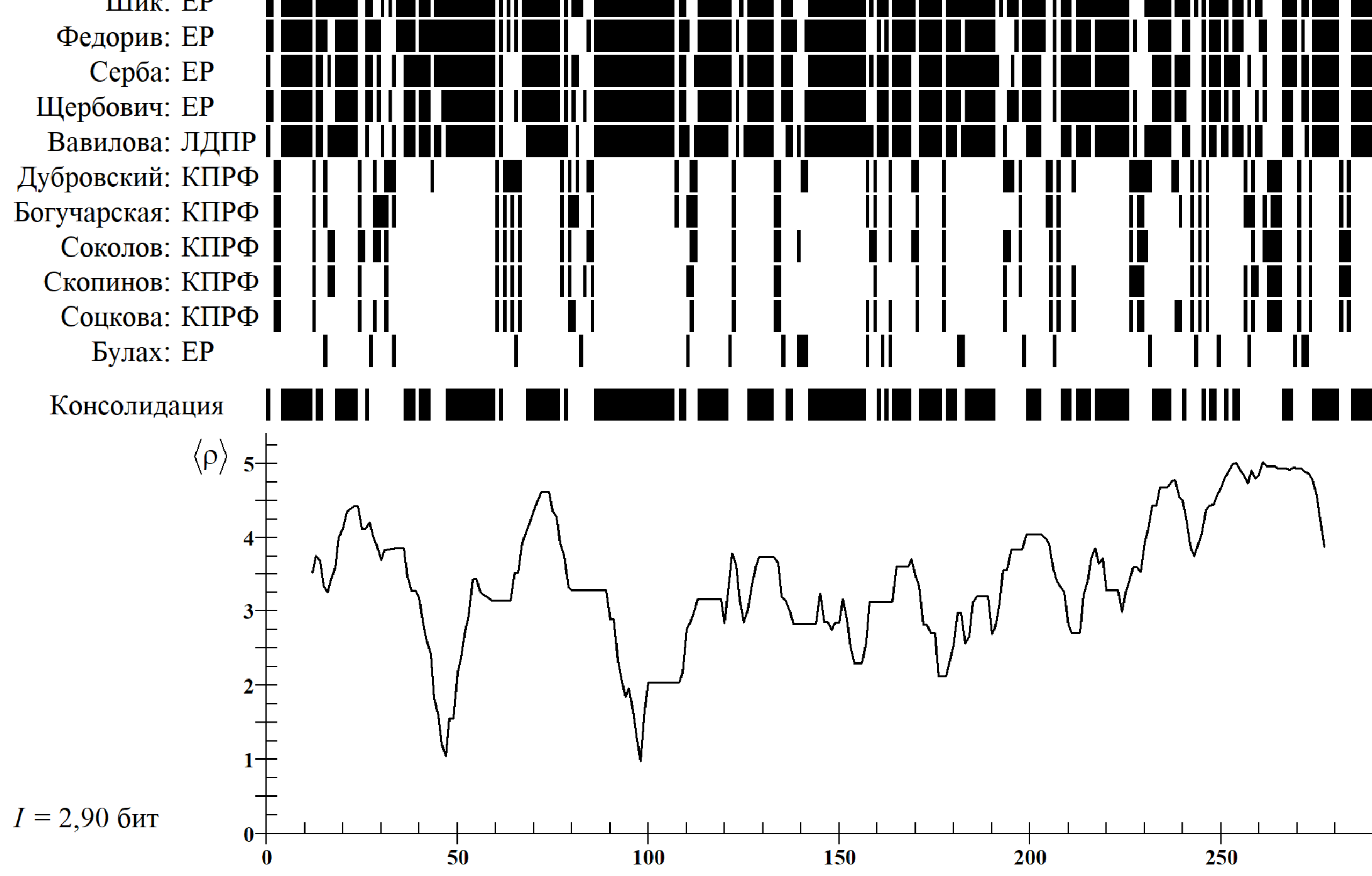

Рис. 7. Стенограмма подсчета голосов для участка 217

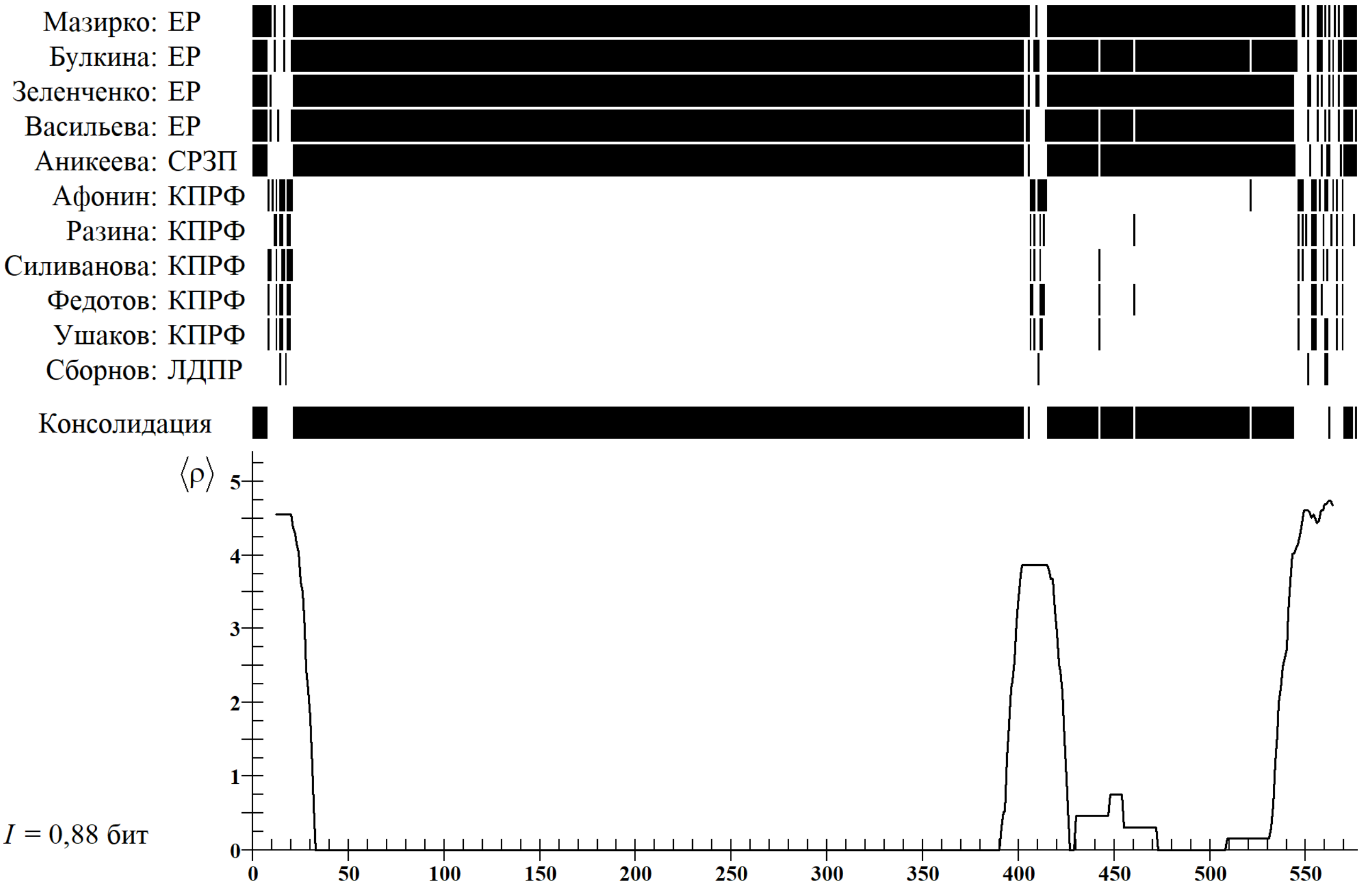

Рис. 8. Стенограмма подсчета голосов для участка 218

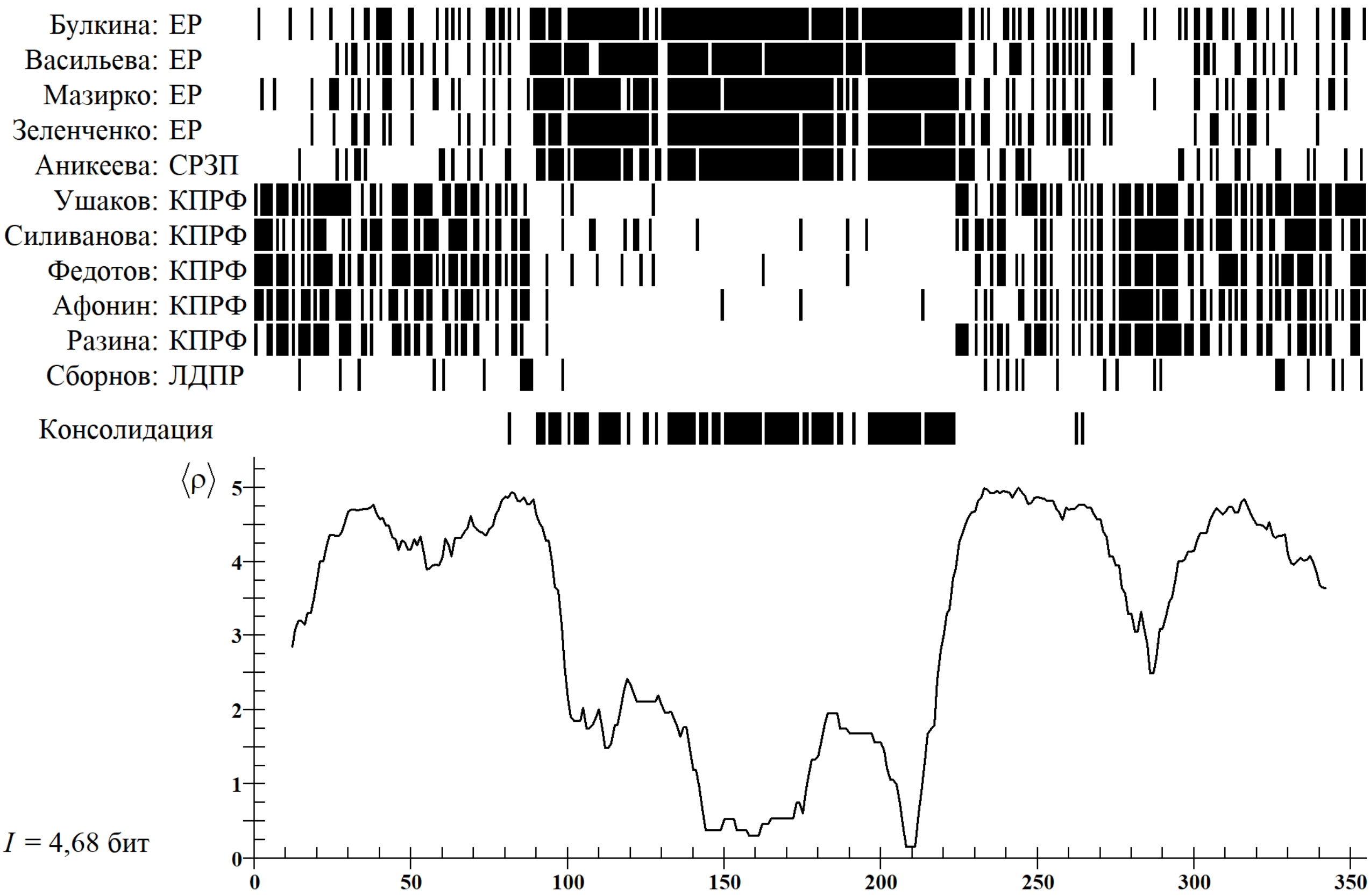

Рис. 9. Стенограмма подсчета голосов для участка 219

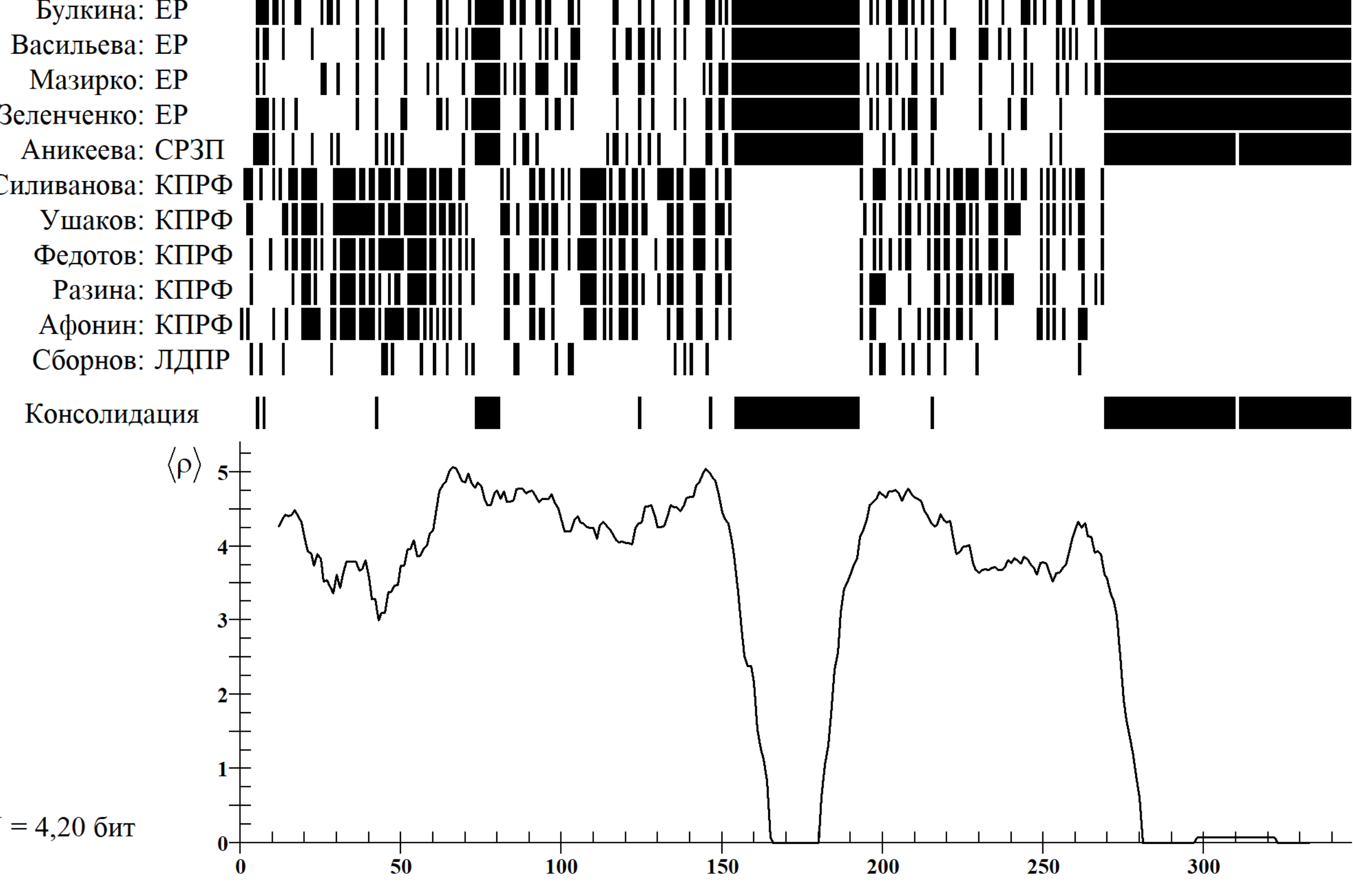

Рис. 10. Стенограмма подсчета голосов для участка 220

Для участка 212 (рис. 2), энтропия весьма высока (выше только на участке 215). Фальсификации здесь были сложными, что маскирует их последствия, в т.ч. повышая энтропию. Дополнительную маскировку обеспечило и умышленное перемешивание бюллетеней. Тем не менее, и для консолидированной стенограммы видны длинные серии голосов «за», и на графике среднего расстояния имеются области, в которых различие бюллетеней заметно падает.

Для участка 213 (рис. 3) энтропия несколько ниже, чем для участка 212, но всё еще недостаточно низка, чтобы можно было говорить о масштабных фальсификациях. Зато они здесь были простыми, что делает их последствия намного более явными, чем на рис. 2. Запись начала подсчета голосов для этого участка отсутствует, из-за чего невозможно утверждать наверняка, имело ли место умышленное перемешивание бюллетеней перед их подсчетом. Однако вид и стенограммы, имеющей множество одинаково закрашенных областей небольшой длины, и графика, несмотря на сильные колебания, не достигающего оси абсцисс, подсказывают положительный ответ на этот вопрос.

Для участка 214 (рис. 4), где фальсификации были простыми и весьма масштабными, а умышленного перемешивания бюллетеней не было, можно видеть и очень низкую энтропию, и столь длинные серии идентичных бюллетеней, что их различие местами падает до нуля.

Для участка 215 (рис. 5), где признаков фальсификаций не зафиксировано, энтропия максимальна. Здесь волеизъявление избирателей столь разнообразно, что график различия бюллетеней почти горизонтален, а при их консолидации лишь 1 избиратель сумел проголосовать «за», угадав тот перечень из 1 провластного и 4 оппозиционных кандидатов, которые победили на данном участке.

Для участка 216 (рис. 6) ситуация является промежуточной между участками 213 и 214 как по масштабам фальсификаций, так и по тщательности перемешивания бюллетеней. Поэтому график, скорее, похож на изображенный на рис. 3, а энтропия здесь почти столь же низка, как и на рис. 4.

Для участка 217 (рис. 7) можно было бы ожидать той же картины, что и на участках 214 и 216, поскольку энтропия столь же низка. Однако по сравнению с рис. 4 и 6 график различия бюллетеней здесь не опускается до нуля из-за очень тщательного перемешивания бюллетеней.

Для участка 218 (рис. 8) в гипертрофированных формах проявили себя все признаки фальсификаций, зафиксированные для участков 214 и 216 (рис. 4 и 6). График различия бюллетеней здесь практически не отлипает от нуля, а энтропия падает ниже 1 бита на бюллетень. К такому результату привел массовый нагон избирателей в 1-й день голосования, результаты которого (равно как и результаты 2-го дня) были фальсифицированы почти полностью.

Для участков 219 и 220 (рис. 9 и 10), картина неоднозначна. Здесь масштаб фальсификаций был сравнительно невелик, что приближает эти участки по энтропии к участку 213 (рис. 3). Однако здесь не было перемешивания, из-за чего с точки зрения консолидации бюллетеней и графиков их различия участок 219, скорее, похож на участок 216 (рис. 6), а участок 220 – на участок 214 (рис. 4).

Определение типа фальсификаций

Ранее была отмечена странность участка 212, бюллетени которого имеют слишком мало отметок по сравнению с другими номерными участками. Официальные итоги голосования на этом участке имеют еще одну странность политологической природы, которая и потребовала типизации фальсификаций.

В каждом из 3 округов по 1 кандидату выдвинула партия ЛДПР, не пользующаяся заметной популярностью в европейской части страны.

В округе №3 кандидат от ЛДПР (Сборнов) закономерно оказался последним по числу голосов на всех участках (обычных и ДЭГ).

В округе №2 кандидатка от ЛДПР (Вавилова), входящая в число провластных кандидатов, была предпоследней на участках 215 и ДЭГ, где не было фальсификаций, но оказалась в пятерке победителей на участках 216 и 217, где фальсификации были.

А вот округе №1 кандидат от ЛДПР (Бахтинов) оказался последним на участках 213 и ДЭГ, предпоследним на участке 214, но удивительным образом 6-м на участке 212. Такой локальный успех слабого кандидата от непопулярной партии невозможен без поддержки со стороны фальсификаторов. Однако они не имели явных причин помогать этому кандидату, поскольку в число провластных кандидатов он не входит. Данное противоречие и говорит о том, что фальсификации здесь были сложными, а не простыми, т.е. осуществлялись не за власть, а против оппозиции.

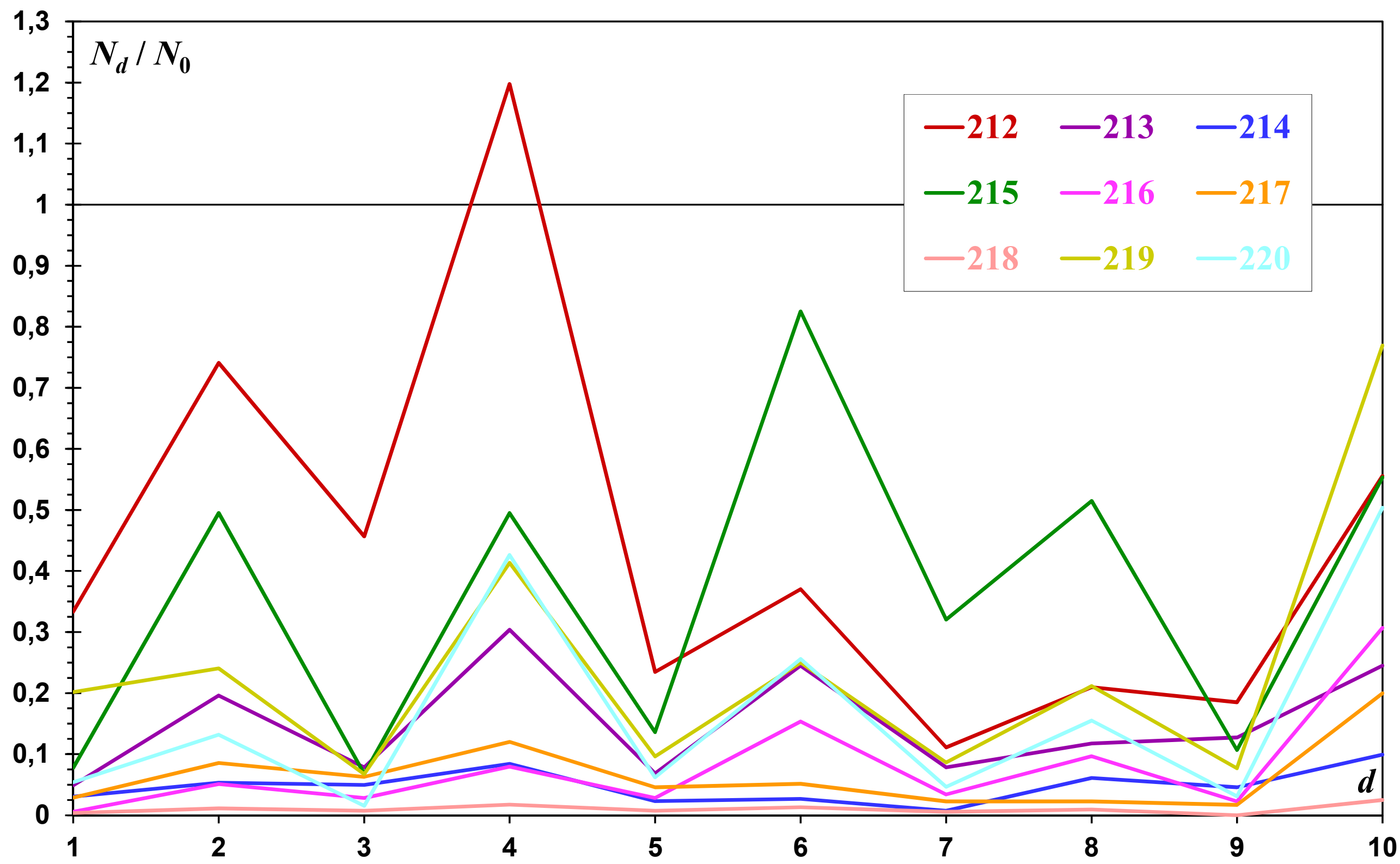

Рис. 11. Отношение количества бюллетеней, находящихся на заданном расстоянии от самого распространенного бюллетеня, к его количеству

Отличить простые фальсификации от сложных без политологического анализа позволяет следующая формальная процедура. Определим самый распространенный вариант заполнения бюллетеня $j = \arg\max\{n_i\}$. Подсчитаем количество всех бюллетеней $N_d = \sum_{\rho(i,j)=d} n_i$, отстоящих от бюллетеня $j$ на расстояние $0 \le d \le 10$ по Хеммингу. Тогда для простых фальсификаций $\arg\max N_d = 0$ (вбрасываются бюллетени, заполненные лишь одним способом), а для сложных фальсификаций $\arg\max N_d > 0$ (вбрасываются бюллетени, заполненные многими способами).

Результат применения этой процедуры иллюстрирует рис. 11. Как можно видеть, только для участка 212 отношение $N_d / N_0$ поднимается над единицей. А второй по высоте максимум графиков уже приходится на участок 215, где фальсификаций не было. Пилообразный вид графиков на рисунке обусловлен тем, что расстояние между бюллетенями с одинаковым числом отметок – всегда четное. А поскольку в большинстве бюллетеней сделаны все 5 допустимых отметок, то закономерно преобладают четные расстояния.

### Строгие методы анализа: Доказательство наличия фальсификаций

Строгие формальные методы электоральной статистики базируются на очевидных или легкопроверяемых предположениях. Это ограничивает применимость таких методов лишь выявлением фальсификаций без определения того, какими были истинные итоги голосования. Зато получаемые результаты обладают доказательной силой.

Исходные данные для анализа

Результаты первичной обработки стенограмм для всех кандидатов и результатов консолидации сведены в табл. 2, 3 и 4 по избирательным округам для всех их участков. Кандидаты в таблицах отсортированы по их суммарному официальному результату на обычных участках.

Таблица 2. Результаты обработки стенограмм для избирательного округа №1

| Кандидат | Партия | Участок 212 | | | | | Участок 213 | | | | | Участок 214 | | | | |
|---|---|---|---|---|---|---|---|---|---|---|---|---|---|---|---|---|
| | | $r_0$ | $s_0$ | $r_1$ | $s_1$ | $m$ | $r_0$ | $s_0$ | $r_1$ | $s_1$ | $m$ | $r_0$ | $s_0$ | $r_1$ | $s_1$ | $m$ |
| Кабанов | ЕР | 162 | 11 | 275 | 14 | 155 | 110 | 10 | 146 | 14 | 91 | 84 | 13 | 304 | 104 | 53 |
| Володичева | ЕР | 185 | 9 | 252 | 12 | 183 | 103 | 10 | 153 | 20 | 91 | 86 | 10 | 302 | 72 | 54 |
| Клюкин | ЕР | 235 | 11 | 202 | 9 | 187 | 118 | 10 | 138 | 11 | 90 | 90 | 15 | 298 | 72 | 61 |
| Григорян | ЕР | 247 | 12 | 190 | 9 | 167 | 113 | 12 | 143 | 16 | 89 | 102 | 19 | 286 | 71 | 51 |
| Лешко | ЕР | 266 | 24 | 171 | 12 | 173 | 117 | 17 | 139 | 10 | 94 | 98 | 15 | 290 | 102 | 65 |
| Яблочкина | КПРФ | 313 | 16 | 124 | 4 | 158 | 179 | 22 | 77 | 8 | 81 | 324 | 163 | 64 | 9 | 59 |
| Бганцов | КПРФ | 329 | 59 | 108 | 5 | 137 | 188 | 32 | 68 | 5 | 82 | 334 | 102 | 54 | 8 | 59 |
| Устинов | КПРФ | 341 | 56 | 96 | 7 | 116 | 198 | 32 | 58 | 4 | 80 | 333 | 109 | 55 | 9 | 53 |
| Слепченко | КПРФ | 353 | 35 | 84 | 4 | 115 | 187 | 32 | 69 | 4 | 89 | 334 | 102 | 54 | 7 | 64 |
| Бахтинов | ЛДПР | 307 | 16 | 130 | 4 | 196 | 222 | 29 | 34 | 3 | 58 | 362 | 194 | 26 | 2 | 37 |
| Шамова | СРЗП | 369 | 41 | 68 | 5 | 102 | 226 | 45 | 30 | 4 | 47 | 356 | 193 | 32 | 5 | 48 |
| Консолидация | | 197 | 9 | 240 | 15 | 175 | 154 | 27 | 102 | 10 | 80 | 126 | 34 | 262 | 71 | 35 |

Таблица 3. Результаты обработки стенограмм для избирательного округа №2

| Кандидат | Партия | Участок 215 | | | | | Участок 216 | | | | | Участок 217 | | | | |
|---|---|---|---|---|---|---|---|---|---|---|---|---|---|---|---|---|
| | | $r_0$ | $s_0$ | $r_1$ | $s_1$ | $m$ | $r_0$ | $s_0$ | $r_1$ | $s_1$ | $m$ | $r_0$ | $s_0$ | $r_1$ | $s_1$ | $M$ |
| Шик | ЕР | 274 | 9 | 199 | 6 | 220 | 109 | 11 | 209 | 31 | 90 | 66 | 5 | 224 | 21 | 88 |
| Федорив | ЕР | 309 | 10 | 164 | 10 | 192 | 119 | 13 | 199 | 31 | 64 | 79 | 5 | 211 | 21 | 86 |
| Серба | ЕР | 316 | 14 | 157 | 4 | 193 | 120 | 9 | 198 | 31 | 82 | 81 | 6 | 209 | 21 | 82 |
| Щербович | ЕР | 323 | 13 | 150 | 6 | 202 | 123 | 13 | 195 | 33 | 60 | 85 | 4 | 205 | 21 | 88 |
| Вавилова | ЛДПР | 356 | 14 | 117 | 4 | 184 | 116 | 13 | 202 | 31 | 75 | 86 | 6 | 204 | 21 | 88 |
| Дубровский | КПРФ | 245 | 8 | 228 | 8 | 220 | 223 | 33 | 95 | 6 | 93 | 226 | 21 | 64 | 6 | 80 |
| Богучарская | КПРФ | 258 | 14 | 215 | 7 | 216 | 234 | 34 | 84 | 6 | 92 | 237 | 26 | 53 | 4 | 76 |
| Соколов | КПРФ | 271 | 9 | 202 | 7 | 208 | 233 | 33 | 85 | 5 | 80 | 237 | 28 | 53 | 5 | 70 |
| Скопинов | КПРФ | 285 | 11 | 188 | 7 | 207 | 232 | 33 | 86 | 5 | 84 | 243 | 28 | 47 | 4 | 68 |
| Соцкова | КПРФ | 298 | 16 | 175 | 6 | 182 | 242 | 33 | 76 | 6 | 84 | 246 | 28 | 44 | 4 | 72 |
| Булах | ЕР | 378 | 22 | 95 | 3 | 148 | 300 | 62 | 18 | 1 | 36 | 265 | 31 | 25 | 3 | 42 |
| Консолидация | | 472 | 353 | 1 | 1 | 2 | 142 | 19 | 176 | 31 | 56 | 115 | 11 | 175 | 21 | 70 |

Таблица 4. Результаты обработки стенограмм для избирательного округа №3

| Кандидат | Партия | Участок 218 | | | | | Участок 219 | | | | | Участок 220 | | | | |
|---|---|---|---|---|---|---|---|---|---|---|---|---|---|---|---|---|
| | | $r_0$ | $s_0$ | $r_1$ | $s_1$ | $m$ | $r_0$ | $s_0$ | $r_1$ | $s_1$ | $m$ | $r_0$ | $s_0$ | $r_1$ | $s_1$ | $m$ |
| Булкина | ЕР | 37 | 5 | 540 | 383 | 30 | 157 | 10 | 198 | 47 | 113 | 152 | 10 | 194 | 78 | 93 |
| Мазирко | ЕР | 32 | 5 | 545 | 385 | 26 | 183 | 13 | 172 | 35 | 100 | 170 | 17 | 176 | 77 | 81 |
| Васильева | ЕР | 43 | 8 | 534 | 383 | 28 | 175 | 26 | 180 | 29 | 98 | 170 | 13 | 176 | 77 | 85 |
| Зеленченко | ЕР | 40 | 11 | 537 | 382 | 24 | 189 | 26 | 166 | 42 | 88 | 176 | 18 | 170 | 77 | 67 |
| Аникеева | СРЗП | 45 | 13 | 532 | 382 | 18 | 199 | 30 | 156 | 32 | 84 | 182 | 21 | 164 | 41 | 69 |
| Ушаков | КПРФ | 558 | 386 | 19 | 3 | 26 | 210 | 96 | 145 | 12 | 106 | 228 | 77 | 118 | 13 | 104 |
| Силиванова | КПРФ | 556 | 385 | 21 | 3 | 30 | 216 | 32 | 139 | 14 | 114 | 227 | 77 | 119 | 8 | 106 |
| Федотов | КПРФ | 557 | 386 | 20 | 3 | 26 | 220 | 40 | 135 | 7 | 118 | 246 | 77 | 100 | 8 | 108 |
| Афонин | КПРФ | 547 | 385 | 30 | 5 | 30 | 238 | 55 | 117 | 11 | 124 | 260 | 82 | 86 | 6 | 91 |
| Разина | КПРФ | 555 | 386 | 22 | 3 | 34 | 241 | 130 | 114 | 8 | 103 | 256 | 77 | 90 | 6 | 100 |
| Сборнов | ЛДПР | 571 | 392 | 6 | 2 | 10 | 327 | 134 | 28 | 4 | 46 | 316 | 84 | 30 | 2 | 52 |
| Консолидация | | 53 | 18 | 524 | 382 | 18 | 251 | 90 | 104 | 17 | 44 | 217 | 53 | 129 | 41 | 19 |

Используемые обозначения: $r_0$ и $r_1$ – число бюллетеней, соответственно не содержащих или содержащих отметку за кандидата, $s_0$ и $s_1$ – максимальная длина серии бюллетеней соответственно без отметки или с отметкой, идущих подряд, $m$ – число переключений между сериями бюллетеней с отметкой и без нее. Сумма $r_0 + r_1 = n$ одинакова для всех кандидатов на участке и равна числу действительных бюллетеней, представленных в стенограмме.

На основе этих величин вычисляются эмпирические вероятности голосования «против» (неголосования «за») $p_0 = q_1 = r_0/n$ и голосования «за» (неголосования «против») $p_1 = q_0 = r_1/n$.

Общее представление о проверке гипотез

Во многих строчках стенограмм на рис. 2-10 отчетливо видно, как области хаотической раскраски, скорее всего, соответствующие голосованию реаль-ных избирателей, прерываются непрерывными областями одного цвета – *аномалиями*, – предположительно возникшими в результате вброса идентично или схоже заполненных бюллетеней. Чтобы объективно оценить это предположение, используется инструментарий проверки *статистических гипотез*.

Формулируются *нулевая* и *альтернативная* гипотезы. Согласно нулевой гипотезе все аномалии являются кажущимися, т.е. возникли естественным путем в силу случайного стечения обстоятельств, тогда как альтернативная гипотеза утверждает, что имело место какое-то неучтенное вмешательство в данные, которое и привело к возникновению аномалий.

На основе информации о природе исследуемых данных и об устройстве рассматриваемых аномалий выбирается или конструируется *статистика* – некоторая числовая характеристика исхода, имеющая известное вероятностное распределение. С его помощью вычисляется *значимость* $\alpha$ нулевой гипотезы – вероятность получения наблюдаемого или даже более аномального значения статистики при условии, что эта гипотеза верна.

Применительно к анализу итогов голосования $\alpha$ определяет вероятность наличия в них аномалий, которые нельзя было бы счесть появившимися случайно. Таким образом, чем ниже значимость, тем с большей степенью уверенности можно утверждать, что именно фальсификации стали причиной аномалий. При этом высокая значимость не позволяет утверждать что-либо о том, имели место фальсификации или нет, она лишь означает, что, если они и были, то обнаружить их последствия не удалось.

В тех случаях, когда значимость не поддается точному вычислению, допустимо ее ограничение сверху посредством указания величины, которую $\alpha$ заведомо не превосходит. Это ослабляет тест, т.е. влечет риск не распознать некоторые аномалии, обусловленные фальсификациями, как таковые, но не влечет риска выдвинуть необоснованные обвинения.

Контекст проверки

Далее описаны и применены два статистических теста, позволяющих выявлять последствия вброса пачек бюллетеней, заполненных не избирателями, волеизъявление каждого из которых индивидуально, а фальсификаторами, стремящимися сместить итоги голосования в пользу определенных кандидатов.

Первый тест, использующий в качестве статистик величины $s_0$ и $s_1$, ориентирован на проверку возможности возникновения естественным путем очень длинных серий идентичных отметок в бюллетенях, идущих подряд, а второй, использующий в качестве статистики величину $m$, – на проверку достаточности общего числа таких серий. Тесты применяются на уровне избирательных участков как к результатам отдельных кандидатов, так и к консолидированным данным бюллетеней.

Такое многообразие аналитических усилий отнюдь не является избыточным, поскольку исследование должно не только убедительно выявлять последствия фальсификаций тогда, когда они имели место, но и быть устойчиво к по-пыткам замаскировать фальсификации с помощью разного рода ухищрений. Кроме того, для дополнительного подтверждения корректности разрабатываемого инструментария и правильности его применения необходим и отрицательный контроль, т.е. ситуации, когда фальсификации выявить не удается как в силу того, что они не имели места, так и в силу того, что фальсификаторы преуспели в маскировке, выведя исследуемый материал за возможности некоторых тестов. Настоящий анализ удовлетворяет всем этим требованиям.

В отличие от большинства тестов, традиционно используемых для проверки статистических гипотез, предлагаемые здесь тесты являются точными, а не асимптотическими, т.е. они применимы к наборам данных любого, а не только очень большого объема.

Тесты базируются на предположениях, что в отсутствии фальсификаций:

1. последовательные бюллетени независимы;
2. вероятности голосования постоянны во времени.

В основе предположения 1 лежат особенности процедуры голосования. Даже если избиратели схожих политических взглядов (родственники, соседи, сослуживцы) голосуют одновременно, их бюллетени случайным образом падают в избирательную урну, где перемешиваются, а при вываливании ее содержимого на стол и собирании в стопку для подсчета они перемешиваются еще раз. Кроме того, консолидация бюллетеней усиливает их независимость, поскольку даже при схожих политических взглядах избиратели могут голосовать не вполне одинаково, что влияет на консолидированные данные.

Благодаря тому, что имеется участок без фальсификаций, предположение 2 допускает прямую проверку. Для ее осуществления описан и применен вспомогательный статистический тест, подтвердивший справедливость этого предположения для анализируемых данных.

Первый тест – на длиннейшую серию успехов

В рамках данного теста рассматривается нулевая гипотеза, гласящая, что в последовательности $n$ испытаний с вероятностями успеха $p$ и неудачи $q=1-p$ серия из не менее чем $s$ успехов, идущих подряд, возникла в силу случайных причин. Значимость $\alpha$ этой гипотезы рассчитывается следующим образом.

Обозначим $u_{n,i,j}$ вероятность того, что в последовательности $n$ испытаний наибольшая длина серии успехов, идущих подряд, равна $i$ при условии, что ровно $j$ последних испытаний тоже завершились успехом. Тогда безусловная вероятность того, что наибольшая длина серии успехов равна $i$, есть $w_{n,i}=\sum_{j=0}^{i}u_{n,i,j}$. Вероятность естественного возникновения серии из не менее чем $s$ успехов $\alpha=\sum_{i=s}^{n}w_{n,i}$.

Величины $u_{n,i,j}$ вычисляются с помощью итерационного процесса, шаг которого соответствует приросту длины последовательности, происходящему при оглашении содержимого очередного бюллетеня: $u_{n+1,i,i} = p \cdot (u_{n,i,i-1} + u_{n,i-1,i-1})$, $u_{n+1,i,j} = p \cdot u_{n,i,j-1}$ для $0 < j < i$ и $u_{n+1,i,0} = q \cdot w_{n,i}$.

Начальное условие $u_{0,0,0} = 1$ соответствует отсутствию уже оглашенных бюллетеней до начала процедуры подсчета, а граничные условия $u_{n,n+1,j} = 0$ и $w_{n,n+1} = 0$ запрещают исходы с невозможными комбинациями индексов.

В данном тесте по каждому набору данных единообразно вычисляются две *частные значимости* $\alpha_0$ и $\alpha_1$. Для первой в качестве успеха рассматривается отсутствие в бюллетене отметки (голосование «против»), а для второй – ее наличие (голосование «за»).

В качестве дополнительной информации вычисляется оценка *общей значимости* $\alpha_* = 1 - (1 - \min\{\alpha_0, \alpha_1\})^2$, представляющая минимум частных значимостей, поправленный на множественность испытаний в предположении возможной зависимости их исходов. Эта величина удобна ее безразличием к тому, осуществлялись ли гипотетические фальсификации за или против кандидата, однако может быть менее чувствительна к ним.

<u>Второй тест – на число переключений между сериями</u>

В рамках данного теста рассматривается нулевая гипотеза, гласящая, что в последовательности $n$ испытаний с вероятностями успеха $p$ и неудачи $q = 1 - p$ общее число переключений между сериями успехов и неудач, идущих подряд, не превысило $m$ в силу случайных причин. Значимость $\tilde{\alpha}$ этой гипотезы рассчитывается следующим образом.

Обозначим $a_{n,i}$ и $b_{n,i}$ вероятности того, что в последовательности $n$ испытаний происходит $i$ переключений между сериями успехов и неудач, идущих подряд, при условии, что последнее испытание последовательности завершилось соответственно успехом или неудачей. Тогда безусловная вероятность того, что имеется $i$ переключений, есть $v_{n,i} = a_{n,i} + b_{n,i}$. Вероятность того, что наблюдаемое число переключений $m$ оказалось мало по естественным причинам, $\tilde{\alpha} = \sum_{i=0}^{m} v_{n,i}$.

Величины $a_{n,i}$ и $b_{n,i}$ вычисляются с помощью итерационного процесса, шаг которого соответствует приросту длины последовательности, происходящему при оглашении содержимого очередного бюллетеня: $a_{n+1,i} = p \cdot (a_{n,i} + b_{n,i-1})$ и $b_{n+1,i} = q \cdot (b_{n,i} + a_{n,i-1})$.

Начальные условия $a_{1,0} = p$ и $b_{1,0} = q$ соответствует ситуации, когда оглашен только первый бюллетень и переключений еще нет, а граничные условия $a_{n,-1} = b_{n,-1} = a_{n,n} = b_{n,n} = 0$ запрещают исходы с невозможными комбинациями индексов.

Для данного теста несущественно, что именно считать успехом, а что неудачей – наличие или отсутствие отметки за кандидата, т.к. подобная инверсия не меняет числа переключений $m$.

Вспомогательный тест – на стационарность потока бюллетеней

В рамках данного теста рассматривается нулевая гипотеза, гласящая, что в последовательности $n$ испытаний с $r$ успехами и $n-r$ неудачами локальные вероятности успеха и неудачи отклоняются от их средних значений $p=r/n$ и $q=1-r/n$ лишь в силу случайных причин. Значимость $\breve{\alpha}$ этой гипотезы можно оценить следующим образом.

Для вычисления локальных вероятностей последовательность испытаний разбивается на $h$ отрезков равной длины (если $n$ не делится нацело на $h$, то исходы граничных испытаний распределяются между смежными отрезками в пропорциональных долях). Коль скоро гипотеза верна, число успехов $k_i$, приходящихся на отрезок $i=1,2,\ldots h$, распределено биномиально с математическим ожиданием $\mu=r/h$ и дисперсией $\sigma^2=\mu\cdot(1-r/n)$. Если аппроксимировать это распределение нормальным с теми же параметрами, то статистика $\sum_{i=1}^{h}(k_i-\mu)^2/\sigma^2$ описывается $\chi^2$-распределением с $h-1$ степенью свободы. Это позволяет получить оценку значимости $\breve{\alpha}$. Выбор того, считается ли успехом голосование за или против, очевидно, не влияет на результат.

Для применения описанного теста необходимо выбрать подходящее число отрезков $h$. С учетом того, что голосование является трехдневным, что избиратели с разными политическими предпочтениями могут склоняться к голосованию в разные дни или даже в разное время дня и что число избирателей варьируется по дням, здесь принято значение $h=12$. Оно позволяет сочетать как достаточное разрешение по времени голосования (в среднем по 4 отрезка на каждый их 3 дней голосования), так и достаточную представительность отрезков (20÷50 бюллетеней для анализируемых данных).

Вспомогательный тест, в отличие от двух основных, является асимптотическим. Это означает, что он точен только при длине отрезка $n/h\to\infty$. А при конечных длинах отрезков получаемая значимость $\breve{\alpha}$ оказывается занижена, причем тем сильнее, чем она меньше. Поэтому ее очень малым значениям можно доверять лишь качественно (как несомненным признакам непостоянства вероятностей), но не количественно. Однако умеренным и высоким значениям $\breve{\alpha}$ можно доверять и количественно как не обнаруживающим непостоянства вероятностей.

Представление и интерпретация результатов

В случае масштабных фальсификаций значимость гипотезы $\alpha$ о естественном возникновении наблюдаемых статистик может быть очень мала. Поэтому удобно использовать *десятичный показатель* значимости $\mathrm{p}\alpha=-\lg\alpha$. Он является магнитудной характеристикой, т.е. его увеличение на 1 означает уменьшение вероятности отсутствия фальсификаций на порядок.

Если нулевая гипотеза верна, то значимость, рассматриваемая как непрерывная случайная величина, равномерно распределена на отрезке $[0;1]$, поэтому при проведении $t$ испытаний она в среднем 1 раз примет значение $\alpha<1/t$. При представлении результатов такие исходы с $\mathrm{p}\alpha > \lg t$ интерпретируются как *подозрительные*, на порядок более редкие исходы с $\mathrm{p}\alpha > 1 + \lg t$ – как *исключительные*, а на два порядка более редкие исходы с $\mathrm{p}\alpha > 2 + \lg t$ – как *невероятные*. Иными словами, подозрительный исход может естественным путем возникать в каждом тесте, исключительный – 1 раз на 10 тестов, невероятный – 1 раз на 100 тестов. Такой упрощенной градации результатов тестирования обычно бывает достаточно для качественных выводов о выявлении или невыявлении фальсификаций.

При анализе итогов с 9 участков, на каждом из которых баллотируются 11 кандидатов, $t = 2 \cdot 9 \cdot 11 = 198$ для частных значимостей первого теста $\alpha_0$ и $\alpha_1$, т.к. вычисляются сразу 2 величины, и $t = 9 \cdot 11 = 99$ для общей значимости первого теста $\alpha_*$, а также для значимостей второго и вспомогательного тестов $\tilde{\alpha}$ и $\breve{\alpha}$.

В каждом тесте наличие фальсификаций на участке наиболее убедительно вскрывает минимальная по всем $c$ кандидатам значимость $\alpha_{\min}$. Однако эту величину сложно интерпретировать, поскольку она уже не является равномерно распределенной. Поправка на множественность испытаний, восстанавливающая равномерность, дает значимость для *объединения кандидатов* $\alpha' = 1 - (1 - \alpha_{\min})^c$. При ее вычислении игнорируется возможность получения низких значимостей сразу несколькими кандидатами, что ослабляет тест. Однако корректный учет такой возможности потребовал бы как-то принимать во внимание сходства и различия разных кандидатов, что трудновыполнимо.

При анализе результатов с 9 участков для объединения кандидатов, равно как и для консолидации бюллетеней, $t = 2 \cdot 9 = 18$ для частных значимостей первого теста $\alpha'_0$ и $\alpha'_1$ и $t = 9$ для общих значимостей первого теста $\alpha'_*$, а также для значимостей второго и вспомогательного тестов $\tilde{\alpha}'$ и $\breve{\alpha}'$.

<u>Результаты применения тестов</u>

Показатели значимостей описанных тестов по всем округам представлены в табл. 5, 6 и 7 для всех кандидатов, а также для результатов их объединения и консолидации бюллетеней.

Для участка 212 (табл. 5), где имели место сложные фальсификации и перемешивание бюллетеней, первый тест работает не очень хорошо. Для него ожидаемо наиболее заметными оказываются длинные серии голосов, поданных против оппозиции. Но вот второй тест лучше видит фальсификации за власть, для которой число переключений между сериями оказывается слишком мало. Для участка в целом объединение кандидатов оказывается более эффективным, чем консолидация бюллетеней.

Для участка 213 (табл. 5), где фальсификации были простыми, но, скорее всего, имело место перемешивание бюллетеней, первый тест не дает однозначных свидетельств фальсификации. Однако второй тест вполне справляется с их обнаружением. Для участка в целом консолидация бюллетеней уже оказывается более эффективной, чем объединение кандидатов, и позволяет обнаружить фальсификации даже с помощью первого теста в части голосования «против».

Для участка 214 (табл. 5), где не было перемешивания, оба теста работают отлично. Примечательно, что фальсифицированные вероятности отсутствия отметки $p_0$ для провластных кандидатов и наличия отметки $p_1$ для оппозиционных кандидатов здесь столь малы, что уже не могут объяснить естественные серии голосов против первых и за вторых.

Таблица 5. Показатели значимостей для участков избирательного округа №1

| Кандидат | Партия | Участок 212 | | | | | Участок 213 | | | | | Участок 214 | | | | |
|---|---|---|---|---|---|---|---|---|---|---|---|---|---|---|---|---|
| | | $p_{\alpha_0}$ | $p_{\alpha_1}$ | $p_{\alpha^*}$ | $p_{\tilde{\alpha}}$ | $p_{\breve{\alpha}}$ | $p_{\alpha_0}$ | $p_{\alpha_1}$ | $p_{\alpha^*}$ | $p_{\tilde{\alpha}}$ | $p_{\breve{\alpha}}$ | $p_{\alpha_0}$ | $p_{\alpha_1}$ | $p_{\alpha^*}$ | $p_{\tilde{\alpha}}$ | $p_{\breve{\alpha}}$ |
| Кабанов | ЕР | **2,3** | 0,7 | **2,0** | **5,1** | **3,9** | 1,5 | 1,4 | 1,2 | **4,7** | **7,2** | **6,2** | **9,2** | **8,9** | **12,5** | **21,8** |
| Володичева | ЕР | 1,0 | 0,7 | 0,7 | **2,5** | **4,9** | 1,8 | **2,5** | **2,2** | **4,1** | **5,6** | **4,1** | **6,0** | **5,7** | **12,6** | **21,2** |
| Клюкин | ЕР | 0,7 | 0,7 | 0,5 | **2,6** | 1,9 | 1,2 | 0,9 | 1,0 | **5,5** | **6,8** | **7,1** | **6,4** | **6,8** | **11,9** | **19,5** |
| Григорян | ЕР | 0,7 | 0,9 | 0,6 | **5,3** | **2,9** | 2,1 | 2,0 | 1,8 | **5,5** | **7,9** | **8,6** | **7,5** | **8,3** | **19,4** | **24,7** |
| Лешко | ЕР | **3,0** | **2,5** | **2,7** | **3,1** | **3,1** | **3,7** | 0,6 | **3,4** | **4,5** | **4,4** | **6,5** | **11,0** | **10,7** | **13,1** | **23,3** |
| Яблочкина | КПРФ | 0,4 | 0,1 | 0,2 | 1,2 | 1,4 | 1,6 | 1,9 | 1,6 | **2,7** | **3,1** | **11,2** | **4,5** | **10,9** | **5,3** | **16,3** |
| Бганцов | КПРФ | **5,3** | 0,6 | **5,0** | 1,7 | **4,7** | **2,5** | 0,7 | **2,2** | 1,5 | **4,2** | **5,0** | **4,3** | **4,7** | **3,1** | **13,4** |
| Устинов | КПРФ | **4,1** | 2,1 | **3,8** | **2,5** | **4,8** | 1,9 | 0,4 | 1,6 | 0,8 | **4,2** | **5,6** | **5,1** | **5,3** | **4,4** | **10,6** |
| Слепченко | КПРФ | 1,4 | 0,4 | 1,1 | 1,3 | **2,5** | **2,6** | 0,2 | **2,3** | 1,0 | **3,0** | **5,0** | **3,5** | **4,7** | **2,4** | **11,5** |
| Бахтинов | ЛДПР | 0,4 | 0,0 | 0,2 | 0,1 | 0,4 | 0,4 | 0,4 | 0,2 | 0,3 | 0,1 | **4,7** | 0,1 | **4,4** | 1,0 | **5,5** |
| Шамова | СРЗП | 1,2 | 1,5 | 1,2 | 0,8 | 1,3 | 1,0 | 1,4 | 1,1 | 0,6 | 1,6 | **6,0** | **2,9** | **5,7** | 0,8 | **9,2** |
| Объединение | | **4,3** | **1,5** | **4,0** | **4,3** | **3,9** | **2,7** | **1,5** | **2,4** | **4,5** | **6,9** | **10,2** | **10,0** | **9,9** | **18,4** | **23,7** |
| Консолидация | | 0,8 | **1,6** | **1,3** | **4,2** | **3,9** | **4,0** | **1,8** | **3,7** | **6,6** | **7,2** | **14,2** | **10,1** | **13,9** | **37,6** | **30,3** |

Таблица 6. Показатели значимостей для участков избирательного округа №2

| Кандидат | Партия | Участок 215 | | | | | Участок 216 | | | | | Участок 217 | | | | |
|---|---|---|---|---|---|---|---|---|---|---|---|---|---|---|---|---|
| | | $p_{\alpha_0}$ | $p_{\alpha_1}$ | $p_{\alpha^*}$ | $p_{\tilde{\alpha}}$ | $p_{\breve{\alpha}}$ | $p_{\alpha_0}$ | $p_{\alpha_1}$ | $p_{\alpha^*}$ | $p_{\tilde{\alpha}}$ | $p_{\breve{\alpha}}$ | $p_{\alpha_0}$ | $p_{\alpha_1}$ | $p_{\alpha^*}$ | $p_{\tilde{\alpha}}$ | $p_{\breve{\alpha}}$ |
| Шик | ЕР | 0,1 | 0,1 | 0,0 | 0,7 | 0,6 | **2,8** | **3,7** | **3,4** | **7,6** | **6,2** | 0,9 | 0,6 | 0,6 | 1,0 | 0,8 |
| Федорив | ЕР | 0,0 | 2,1 | 1,8 | 1,5 | 0,1 | **3,3** | **4,3** | **4,0** | **18,9** | **8,7** | 0,6 | 1,0 | 0,8 | **2,7** | 1,0 |
| Серба | ЕР | 0,4 | 0,0 | 0,2 | 1,0 | 0,1 | 1,5 | **4,3** | **4,0** | **12,2** | **8,1** | 1,0 | 1,1 | 0,8 | **3,7** | 1,4 |
| Щербович | ЕР | 0,2 | 0,6 | 0,3 | 0,4 | 2,0 | **3,1** | **5,0** | **4,7** | **22,0** | **10,2** | 0,1 | 1,3 | 1,0 | **3,3** | 1,8 |
| Вавилова | ЛДПР | 0,0 | 0,1 | 0,0 | 0,1 | 0,3 | **3,4** | **4,1** | **3,8** | **13,8** | **7,7** | 0,9 | 1,3 | 1,0 | **3,4** | **2,4** |
| Дубровский | КПРФ | 0,2 | 0,3 | 0,1 | 1,1 | 1,6 | **3,2** | 0,8 | **2,9** | **4,5** | **6,4** | 0,6 | 1,6 | 1,3 | 1,6 | 0,3 |
| Богучарская | КПРФ | 1,4 | 0,2 | 1,1 | 1,3 | 1,2 | **2,6** | 1,1 | **2,3** | **2,9** | **5,3** | 0,6 | 0,6 | 0,4 | 0,8 | 0,4 |
| Соколов | КПРФ | 0,1 | 0,3 | 0,1 | 1,7 | 0,3 | **2,6** | 0,6 | **2,3** | **5,1** | **6,5** | 0,8 | 1,3 | 1,0 | 1,3 | 0,2 |
| Скопинов | КПРФ | 0,3 | 0,4 | 0,2 | 1,3 | 1,0 | **2,6** | 0,5 | **2,3** | **4,6** | **5,7** | 0,6 | 0,8 | 0,5 | 0,8 | 1,1 |
| Соцкова | КПРФ | 1,0 | 0,3 | 0,7 | **3,2** | 1,8 | 2,1 | 1,4 | 1,8 | **2,9** | **5,2** | 0,5 | 0,9 | 0,6 | 0,4 | 0,5 |
| Булах | ЕР | 0,3 | 0,0 | 0,1 | 0,4 | 0,9 | 0,4 | 0,0 | 0,2 | 0,2 | 1,0 | 0,1 | 0,8 | 0,5 | 0,4 | 0,2 |
| Объединение | | 0,5 | 1,1 | 0,8 | **2,2** | **1,0** | **2,4** | **4,0** | **3,7** | **21,0** | **9,2** | 0,2 | 0,6 | 0,4 | **2,7** | **1,4** |
| Консолидация | | 0,2 | 0,2 | 0,1 | 0,1 | 0,4 | **4,4** | **5,9** | **5,6** | **29,9** | **12,7** | **2,2** | **2,6** | **2,3** | **14,2** | **3,6** |

Для участка 215 (табл. 6), где фальсификаций не было, имеется один исключительный показатель во втором тесте. По всей видимости, предположение о независимости последовательных бюллетеней выполняется неидеально из-за неполного их перемешивания в урне. Такая возможность изредка может сказываться на результатах тестов, поэтому в пограничных ситуациях следует проявлять известную осторожность при интерпретации результатов. Однако речь идет именно о мелкомасштабных флуктуациях потока голосов, поскольку вспомогательный тест подтверждает его общую стационарность для всех кандидатов на этом участке.

Для участка 216 (табл. 6), где по сравнению с участком 213 перемешивание, возможно, было менее тщательным, его оказывается недостаточно, чтобы скрыть последствия фальсификаций. Наиболее заметны они во втором тесте, но и в первом тоже видны при голосовании за власть.

Для участка 217 (табл. 6), где перемешивание бюллетеней было тщательным, для отдельных кандидатов фальсификации почти не видны, но их хорошо видно для консолидированных данных. Бюллетени перемешаны достаточно хорошо, чтобы обеспечить высокую стационарность потока голосов для каждого кандидата, но не для консолидированных данных.

Для участков 218-220 (табл. 7), где не было умышленного перемешивания, а фальсификации были простыми, они уверенно выявляются всеми тестами. Здесь, как и в случае участка 214, вброшено столь много бюллетеней с отметками за провластных кандидатов, что вероятность отсутствия отметки за них оказывается слишком мала, чтобы объяснить даже естественные серии голосов против. Более экзотичную картину демонстрирует сравнение участка 218, где фальсифицированы были очень масштабны, с участками 219 и 220, где фальсификации были сравнительно умеренными. На участке 218 второй тест не обнаруживает фальсификаций в отношении оппозиционных кандидатов. Из-за экстремально низкой эмпирической вероятности голосования за них здесь даже очень малое число переключений между сериями перестает быть аномальным.

Таблица 7. Показатели значимостей для участков избирательного округа №3

| Кандидат | Партия | Участок 218 | | | | | Участок 219 | | | | | Участок 220 | | | | |
|---|---|---|---|---|---|---|---|---|---|---|---|---|---|---|---|---|
| | | $p\alpha_0$ | $p\alpha_1$ | $p\alpha_*$ | $p\tilde{\alpha}$ | $p\breve{\alpha}$ | $p\alpha_0$ | $p\alpha_1$ | $p\alpha_*$ | $p\tilde{\alpha}$ | $p\breve{\alpha}$ | $p\alpha_0$ | $p\alpha_1$ | $p\alpha_*$ | $p\tilde{\alpha}$ | $p\breve{\alpha}$ |
| Булкина | ЕР | 3,2 | 9,9 | 9,6 | 4,4 | 17,3 | 1,3 | 9,8 | 9,5 | 10,3 | 20,5 | 1,3 | 17,5 | 17,2 | 15,9 | 20,6 |
| Мазирко | ЕР | 3,5 | 8,5 | 8,2 | 3,8 | 18,3 | 1,5 | 8,8 | 8,5 | 16,0 | 22,5 | 3,0 | 20,5 | 20,2 | 23,3 | 26,7 |
| Васильева | ЕР | 6,3 | 11,7 | 11,4 | 6,6 | 26,9 | 5,8 | 6,3 | 6,0 | 17,0 | 24,8 | 1,8 | 20,5 | 20,2 | 21,3 | 22,0 |
| Зеленченко | ЕР | 10,0 | 10,8 | 10,5 | 6,7 | 25,0 | 4,9 | 11,6 | 11,3 | 20,9 | 29,6 | 3,1 | 21,6 | 21,3 | 31,2 | 25,3 |
| Аникеева | СРЗП | 11,7 | 12,3 | 12,0 | 10,4 | 26,0 | 5,4 | 9,2 | 8,9 | 21,2 | 26,5 | 3,7 | 11,1 | 10,8 | 29,4 | 25,1 |
| Ушаков | КПРФ | 4,8 | 1,7 | 4,4 | 0,9 | 7,4 | 19,9 | 2,4 | 19,6 | 10,9 | 24,5 | 12,0 | 3,7 | 11,7 | 6,6 | 16,1 |
| Силиванова | КПРФ | 5,3 | 1,6 | 5,0 | 0,9 | 9,9 | 4,8 | 3,4 | 4,5 | 7,8 | 13,8 | 12,1 | 1,4 | 11,8 | 6,3 | 13,5 |
| Федотов | КПРФ | 5,0 | 1,6 | 4,7 | 1,1 | 6,2 | 6,2 | 0,7 | 5,9 | 6,3 | 18,1 | 9,5 | 1,9 | 9,2 | 3,1 | 13,8 |
| Афонин | КПРФ | 7,9 | 3,7 | 7,6 | 2,6 | 13,9 | 7,6 | 2,9 | 7,3 | 3,0 | 17,4 | 8,4 | 1,2 | 8,1 | 3,7 | 12,7 |
| Разина | КПРФ | 5,6 | 1,5 | 5,3 | 0,7 | 11,0 | 20,0 | 1,6 | 19,7 | 6,4 | 17,0 | 8,2 | 1,1 | 7,9 | 2,9 | 9,0 |
| Сборнов | ЛДПР | 1,3 | 1,2 | 1,0 | 0,3 | 1,6 | 3,5 | 1,9 | 3,2 | 0,5 | 3,1 | 1,9 | 0,0 | 1,6 | 0,4 | 1,3 |
| Объединение | | 10,7 | 11,3 | 11,0 | 9,4 | 25,9 | 19,0 | 10,6 | 18,7 | 20,2 | 28,6 | 11,1 | 20,6 | 20,3 | 30,2 | 25,7 |
| Консолидация | | 16,0 | 14,7 | 15,7 | 13,4 | 32,8 | 11,7 | 6,7 | 11,4 | 23,0 | 35,4 | 8,7 | 15,3 | 15,0 | 51,0 | 39,7 |

Общие итоги тестирования

Тесты на длиннейшую серию и на число переключений между сериями позволяют с полной уверенностью выявлять вброс (подмену) пачек бюллетеней в случае не очень тщательного их перемешивания. Однако в случае тщательного перемешивания бюллетеней эффективность сохраняет лишь тест на число переключений, причем применяемый только к консолидированным бюллетеням.

По сравнению с тестом на длиннейшую серию, тест на число серий чувствительнее к фальсификациям и устойчивее к умеренному перемешиванию бюллетеней и к мелким ошибкам в стенограмме. Но одновременно он и требовательнее к независимости последовательных бюллетеней.

При рассмотрении участка в целом консолидация бюллетеня оказывается эффектнее объединения кандидатов в случае простых фальсификаций, но менее эффективна в случае сложных. Кроме того, консолидация бюллетеней много лучше противостоит их перемешиванию, чем объединение кандидатов.

В отсутствии фальсификаций или при тщательном перемешивании бюллетеней поток голосов демонстрирует высокую стационарность. И хотя вспомогательный тест не дает дополнительных свидетельств фальсификаций, он позволяет удовлетворительно воспроизвести качественные выводы основных тестов.

### Полустрогие методы анализа: Простое выявление фальсификаций

Как и строгие формальные методы электоральной статистики, ее полустрогие методы не используют никаких нетривиальных предположений. Однако такие методы позволяют либо делать лишь качественные выводы о наличии фальсификаций, либо, если и характеризуют их количественно, то получаемые при этом величины сами по себе интерпретировать затруднительно.

Наиболее известным примером ситуации, позволяющей сразу сделать однозначный качественный вывод, является жадное голосование [1], связанное с использованием на избирательном участке с неабсолютной явкой всех бюллетеней, полученных участковой комиссией. Жадное голосование возникает в результате стремления фальсификаторов достичь завышенных целевых показателей явки, бездумно выгребая подчистую все полученные бюллетени. Однако если хотя бы один бюллетень остается неиспользованным, то на этой основе уже нельзя ничего утверждать о наличии фальсификаций.

Пример эффективной количественной характеристики избирательных участков дает удельная информационная энтропия бюллетеней, рассмотренная выше. Ее нормальное значение, которое должно получаться в отсутствии фальсификаций, зависит от числа распределяемых мандатов и соревнующихся кандидатов, а также от конкурентности выборов. Поэтому априори можно утверждать лишь то, что фальсификации уменьшают энтропию. Однако зная общее положение дел для набора участков, сходных по указанным параметрам, апостериори можно построить приблизительную шкалу энтропии. Применительно к рассматриваемым выборам при $I < 1$ фальсификации следует трактовать как чудовищные (участок 218), при $I < 3$ – как сильные (участки 214, 216 и 217), при $I < 5$ – как умеренные (участки 213, 218 и 219), а при бо́льших значениях $I$ – как отсутствующие (участок 215) или сложные (участок 212).

Таким образом, наличие опорных данных позволяет использовать информационную энтропию как простой качественный индикатор простых фальси-фикаций, нечувствительный к перемешиванию. Однако желательно иметь аналогичный индикатор, способный хотя бы отчасти справляться и со сложными фальсификациями. В основу индикатора такого рода, описываемого далее, положено то обстоятельство, что распределения по длине серий бюллетеней, содержащих или не содержащих отметку за кандидата, являются зависимыми. Однако параметры этих распределений возможно определить независимым образом и сопоставить друг с другом.

Уже начавшаяся серия бюллетеней при вероятностях ее продолжения $p$ (голосование) и завершения $q$ (неголосование) будет иметь длину $l$ с вероятностью $p^{l-1}q$. Чтобы не возникала специальная ситуация прерывания серии из-за исчерпания последовательности бюллетеней, при анализе замкнем стенограмму в кольцо, расположив ее первый бюллетень после последнего. При этом средняя длина серий голосов $\langle l \rangle = 1/q$, а ее средний квадрат $\langle l^2 \rangle = (1+p)/q^2$, откуда $p = (z-1)/(z+1)$, где $z = \langle l^2 \rangle / \langle l \rangle$. Принципиально важно, что вероятности голосования против кандидата $p_0$ и за него $p_1$ здесь рассчитываются на основе непересекающихся наборов данных, что позволяет проверить выполнение нормировки $p_0 + p_1 = 1$.

Результаты анализа по всем кандидатам для всех участков представлены на рис. 12. Как можно видеть, только для участка 215 точки лежат вблизи линии нормировки (величина отклонения от нее, усредненная по всем кандидатам, дана на врезке). А уже точки участков 212 (сложные фальсификации, пе-

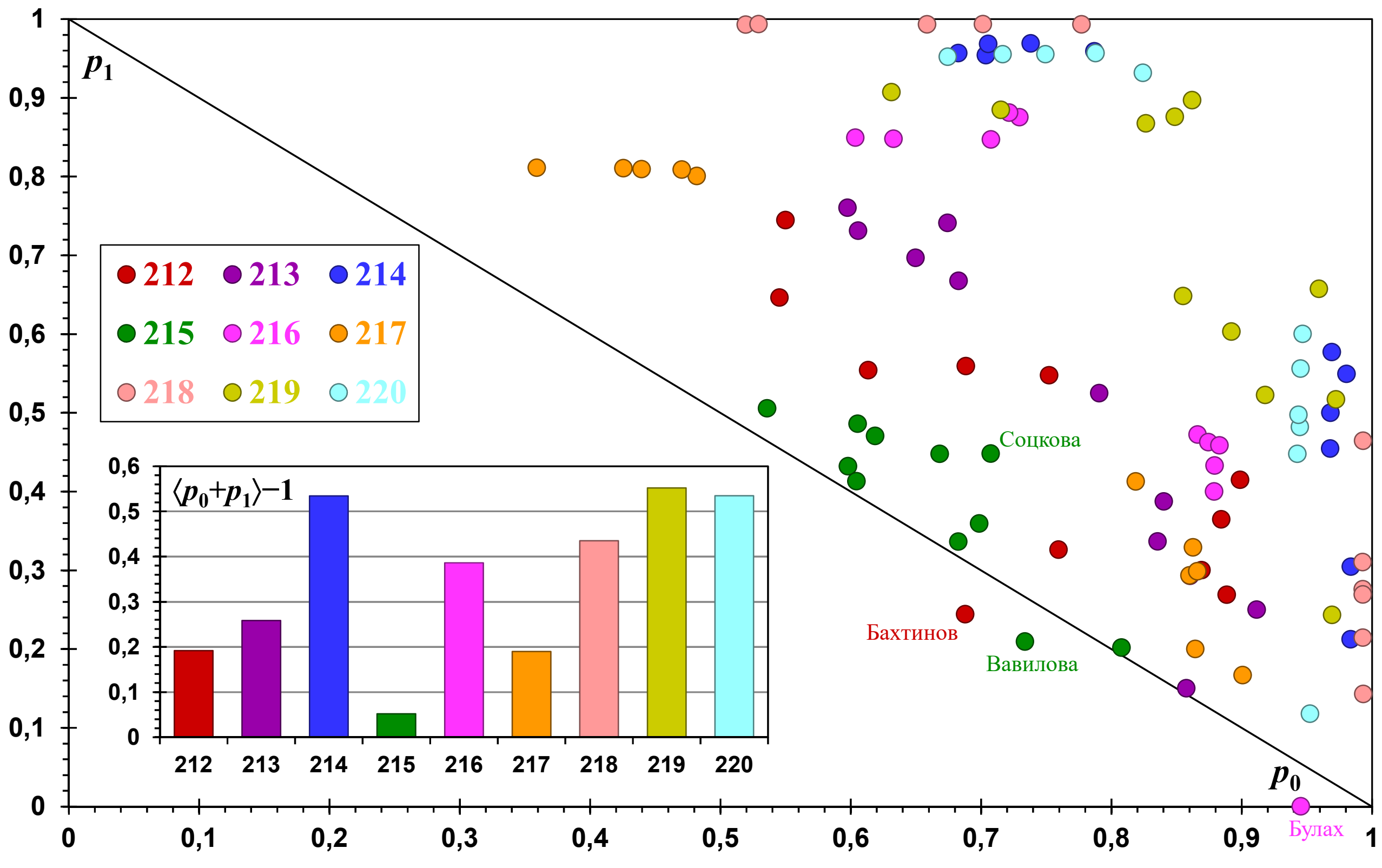

Рис. 12. Проверка нормировки вероятностей голосования ремешивание) и 217 (простые фальсификации, тщательное перемешивание) заметно отклоняются от нее. Для участка 213 (умеренные фальсификации, перемешивание) отклонение еще сильнее, а для прочих участков оно уже тотально.

### Нестрогие методы анализа: Реконструкция итогов голосования

Любая реконструкция представляет собой некорректную задачу, поиск решения которой существенным образом опирается на априорное представление о том, как это решение должно быть устроено. В случае электоральной реконструкции такое представление определяется используемой моделью поведения избирателей. Ее строгое обоснование в общем случае может оказаться невозможным [5], более того, даже ее проверка для конкретной ситуации обычно представляет собой сложную проблему.

Модель обычно выбирается из соображений эффективности, а не доказанности, чем обеспечиваются, с одной стороны, б*о*льшие возможности нестрогих методов электоральной статистики [5,6,7], а с другой – их неспособность отличить гипотетическое несовершенство модели от последствий фальсификаций. Однако, если с помощью строгих или полустрогих методов уже надежно установлен сам факт фальсификаций, то разумно именно на их счет относить любые отклонения данных от модели.

Фальсификация итогов голосования имеет своей задачей искажение лишь общего отношения избирателей к баллотирующимся кандидатам. Однако на многомандатных выборах содержимое отдельных бюллетеней показывает еще и то, как к кандидатам относится каждый отдельный избиратель. Отмечая в бюллетене одновременно нескольких кандидатов, он тем самым констатирует их сходство в смысле соответствия его личным политическим предпочтениям. Поэтому взаимное позиционирование кандидатов столь же информативно, как и их индивидуальная популярность. И фальсификаторам вряд ли по силам исказить оба этих канала информации согласованным образом.

Модель поведения избирателей

Пусть всего имеется $n$ действительных бюллетеней, в $n_i$ из которых отмечен кандидат $i = 1,2,\ldots c$, а в $n_{ij}$ бюллетенях одновременно отмечены два различных кандидата $i$ и $j$. Модельное предположение аппроксимирует взаимосвязь между этими числами формулой $n_{ij}n = \gamma n_i n_j \left(1 + \varepsilon \vec{e}_i \vec{e}_j\right)$, где $\gamma \sim 1$, единичные вектора $\vec{e}_i$ и $\vec{e}_j$ характеризуют политические ориентации кандидатов $i$ и $j$, а параметр $0 \le \varepsilon \le 1$ – политизацию избирателей. Если она высока, то кандидаты с близкими ориентациями чаще получают поддержку одних и тех же избирателей, а если низка, то одновременная поддержка таких кандидатов обуславливается лишь их индивидуальной популярностью, но не политическим сходством.

Кроме того, в рамках данной реконструкции будем считать, что все возможные политические ориентации представлены в равной степени, т.е. $\sum_{i=1}^{c} \vec{e}_i = 0$.

Отношение $p_{ij} = n_{ij} / n_j$ определяет условную вероятность голосования за кандидата $i$ для избирателя, проголосовавшего за кандидата $j$. Сумма этих вероятностей $s_i = \sum_{j \neq i} p_{ij} = kr_i$, где $r_i = n_i / n$ – доля голосов, поданных за кандидата $i$, а коэффициент пропорциональности $k = \gamma(c - 1 - \varepsilon)$.

Если фальсификаций нет, то в координатах $(r, s)$ точки, представляющие кандидатов, лежат на прямой, проходящей через начало координат. А фальсификация итогов в пользу некоторого списка сходно ориентированных кандидатов проявляется как усиление политизации избирателей, но только в отношении этих кандидатов. Иначе говоря, для них происходит увеличение параметра $\varepsilon$, что ведет к уменьшению коэффициента $k$. При этом точки смещаются вправо–вниз относительно упомянутой прямой.

Важно, что направление смещения известно априори. Это позволяет, даже не прибегая к политологическому анализу, понять, в пользу каких именно кандидатов осуществлены фальсификации. Тем не менее, рассмотрение политической принадлежности кандидатов здесь не будет лишним, поскольку облегчает и представление, и понимание результатов.

Политический контекст

В рассматриваемых выборах участвовали кандидаты от 4 (из 5 существующих) привилегированных политических партий: КПРФ, ЕР, СРЗП и ЛДПР.

В каждом из 3 избирательных округов из 11 кандидатов, претендовавших на 5 мест в Совете, можно выделить 5 оппозиционных и 5 провластных кандидатов, имевших более-менее существенную поддержку избирателей, и еще 1 кандидата от политического болота, который на этих выборах оставался статистом. Раскладка партий, выдвинувших кандидатов, по указанным группам представлена в табл. 8. Основная политическая конкуренция развернулась между кандидатами от ЕР (власть) и КПРФ (оппозиция). А кандидаты от СРЗП и ЛДПР либо, не сумев заручиться поддержкой ни избирателей, ни фальсификаторов, оказывались аутсайдерами, либо де факто примыкали к власти или оппозиции (впрочем, отнесение кандидатки от СРЗП в округе №1 именно к оппозиции, а не к болоту является дискуссионным) там, где сильнейшие партии не сумели выдвинуть сразу 5 перспективных кандидатов.

Таблица 8. Политическая ориентация кандидатов по округам

| Округ | Оппозиция | Болото | Власть |
|---|---|---|---|
| №1 | 4 КПРФ<br>1 СРЗП | 1 ЛДПР | 5 ЕР |
| №2 | 5 КПРФ | 1 ЕР | 4 ЕР<br>1 ЛДПР |
| №3 | 5 КПРФ | 1 ЛДПР | 4 ЕР<br>1 СРЗП |

На рис. 13 приведены примеры различных ситуаций, возникших на избирательных участках. На участке 215 фальсификаций не было. Участки 213 и 216 дают примеры простых фальсификаций (в пользу власти), а участок 212 – сложных (против оппозиции). Участок без фальсификаций подтверждает общую обоснованность предложенной модели поведения избирателей, а участки с фальсификациями – корректность качественного описания их последствий,

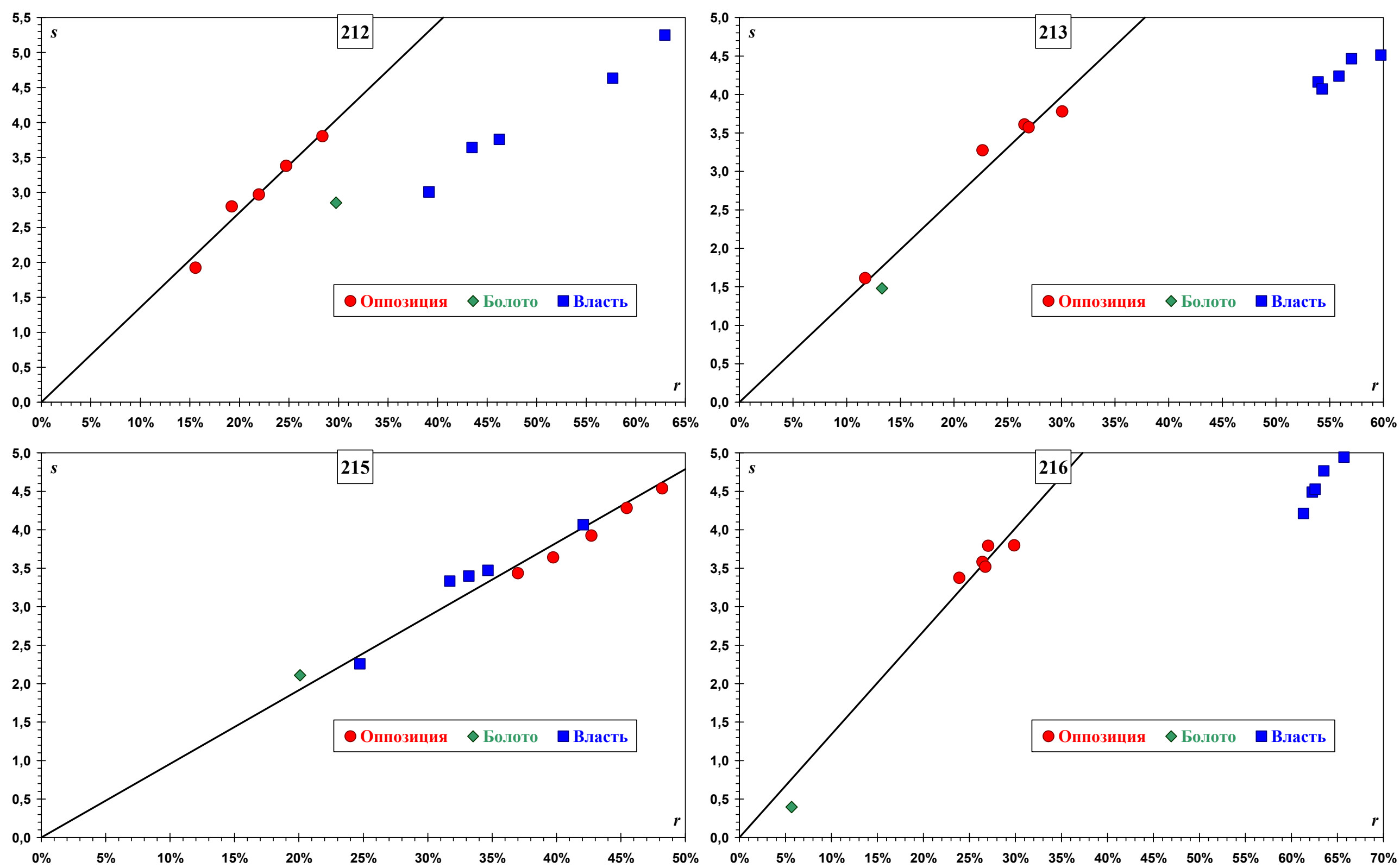

Рис. 13. Сумма условных вероятностей голосования за кандидата в зависимости от его результата – исходные данные предсказанных моделью. Достоверные точки ложатся на регрессионную прямую (при расчете ее углового коэффициента учитывались 5 точек для участка 212, 6 точек – для участков 213 и 216 и все 11 точек – для участка 215), а фальсифицированные отклоняются вправо. Групповая принадлежность кандидатов на рисунке обозначена исключительно для наглядности, поскольку и без нее можно видеть, какие точки достоверны, а какие фальсифицированы.

На рассматриваемых выборах болото в политическом смысле ближе ко власти, чем к оппозиции. Из-за этого сделанное выше предположение о равной представленности политических ориентаций может выполняться не вполне точно, что чревато занижением реконструированного результата оппозиции, получаемого далее. Поэтому его следует рассматривать как минимальную оценку.

Реконструкция для простых фальсификаций

Простые фальсификации трактуются как вброс $\delta$ бюллетеней, в которых отмечены все провластные кандидаты (и только они). При этом реконструкция количества и содержимого подлинных бюллетеней проводится по формулам $\hat{n} = n - \delta$, $\hat{n}_i = n_i - f_i\delta$ и $\hat{n}_{ij} = n_{ij} - f_i f_j \delta$, где $f_i$ – флаг фальсификаций в пользу кандидата $i$, т.е. $f_i = 1$ для провластных кандидатов и $f_i = 0$ для прочих.

Истинные условные вероятности $\hat{p}_{ij} = \hat{n}_{ij} / \hat{n}_j$, их суммы $\hat{s}_i = \sum_{j \neq i} \hat{p}_{ij}$ и доли голосов $\hat{r}_i = \hat{n}_i / \hat{n}$ рассчитываются на основе реконструированных значений $\hat{n}$, $\hat{n}_i$ и $\hat{n}_{ij}$. В отсутствие фальсификаций суммы вероятностей и доли голосов взаимно пропорциональны. Регрессионные угловые коэффициенты для зависимостей $\hat{s} = k\hat{r}$ и $\hat{r} = k'\hat{s}$ равны $k = \sum_i \hat{s}_i \hat{r}_i \Big/ \sum_i \hat{r}_i^2$ и $k' = \sum_i \hat{s}_i \hat{r}_i \Big/ \sum_i \hat{s}_i^2$ соответственно. Оптимальное значение $\delta$ находится максимизацией произведения $kk'$, которое тем ближе к 1, чем лучше точки ложатся на прямую.

Бюллетени дискретны, поэтому при определении объема вброса не нужна высокая точность. И чтобы не решать численно чрезвычайно громоздкое нелинейное уравнение, оптимизация выполняется простым сканированием диапазона значений $0 \le \delta < n$ с небольшим шагом, достаточным, чтобы избежать последствий ошибок округления (здесь использована дискретизация до 0,01).

Поскольку на рассматриваемых выборах фальсификации осуществлялись не вбросом бюллетеней, а их подменой, завершающим шагом реконструкции становится масштабирование реконструированных значений на полное число бюллетеней $\hat{n}_i \to \hat{n}_i / (1 - \delta/n)$. Для участка 213 выполняется еще и дополнительное масштабирование с коэффициентом 273/256, компенсирующее неполноту доступной стенограммы подведения итогов.

Реконструкция для сложных фальсификаций

Сложные фальсификации рассматриваются как комбинация вброса $\delta$ бюллетеней, в которых отмечены все провластные кандидаты (и только они), со вбросом $\delta_0$ бюллетеней, в которых случайным образом отмечены $b \le 5$ из $a = 6$ неоппозиционных кандидатов (и только они). При этом реконструкция количества и содержимого подлинных бюллетеней проводится по формулам $\hat{n} = n - \delta - \delta_0$, $\hat{n}_i = n_i - f_i\delta - g_i\delta_1$ и $\hat{n}_{ij} = n_{ij} - f_i f_j \delta - g_i g_j \delta_2$, где $f_i$ и $g_i$ – флаги фальсификаций в пользу кандидата $i$ по одной и другой схеме, т.е. $f_i = 1$ для провластных кандидатов и $f_i = 0$ для прочих, а $g_i = 0$ для оппозиционных кандидатов и $g_i = 1$ для прочих. Влияние фальсификаций против оппозиции на содержимое бюллетеней слабее, чем на их количество: $\delta_1/\delta_0 = b/a$ и $\delta_2/\delta_1 = (b-1)/(a-1)$.

Здесь необходимо максимизировать произведение $kk'$ уже не по одной переменной, а сразу по трем, в число которых кроме $\delta$ входят любые две величины из набора $\delta_0$, $\delta_1$, $\delta_2$ и $b$. Проводить такую оптимизацию сканированием оказывается крайне затратно, поэтому она выполняется приближенно и стадийно, что позволяет сканировать только величины $\delta$ и $\delta_1$.

Для выполнения приближения величины $\delta_0$, $\delta_1$ и $\delta_2$ сначала полагаются взаимно независимыми, и рассматривается пропорциональность $\hat{s} = k\hat{r}$. Если представить сумму условных вероятностей в виде $\hat{s}_i = y_i - x_i\delta_2$, где $y_i = \sum_{j \ne i} (n_{ij} - f_i f_j) / \hat{n}_j$ и $x_i = \sum_{j \ne i} g_i g_j / \hat{n}_j$, то регрессионная модель принимает вид $\sum_i (y_i - x_i\delta_2 - k\hat{r}_i)^2 \to \min$. В долю голосов $\hat{r}$ в виде единой комбинации входят величина $\delta_0$ и коэффициент $k$, что делает невозможным их одновре-менное определение из такой модели. Поэтому на данном шаге принимается произвольное значение $\delta_0 = 0$ и из системы линейных уравнений

$$\begin{cases} k\sum_i \hat{r}_i^2 + \delta_2 \sum_i x_i \hat{r}_i = \sum_i \hat{r}_i y_i \\ k\sum_i x_i \hat{r}_i + \delta_2 \sum_i x_i^2 = \sum_i x_i y_i \end{cases}$$

находится только значение $\delta_2$. Затем на его основе вычисляются параметр $b = 1 + (a-1) \cdot \delta_2 / \delta_1$ и величина $\delta_0 = a/b \cdot \delta_1$, что позволяет пересчитать $\hat{s}_i$ и $\hat{r}_i$, сводя задачу к уже решенной для простых фальсификаций.

Масштабирование реконструированных значений на полное число бюллетеней выполняется по формуле $\hat{n}_i \to \hat{n}_i / (1 - (\delta + \delta_0)/n)$.

Результаты реконструкции

На рис. 14 для пары избирательных участков из каждого округа приведены примеры результатов реконструкции. Здесь, в отличие от рис. 13, кандидаты от власти и от оппозиции перемешаны по результату, что говорит о конкурентном характере выборов. А болото, фактически устранившееся от борьбы, закономерно показывает посредственные результаты.

Параметры реконструкции по всем участкам сведены в табл. 9. Для участка 215 сохранены исходные данные, для участка 212 выполнена реконструкция сложной фальсификации (ее параметры: $\delta_0 = 143,3$, $\delta_1 = 87,2$, $\delta_2 = 46,3$ и $b = 3,653$), а для остальных участков – простых фальсификаций. Наихудшим качество реконструкции оказывается для участков 214 и 218, продемонстрировавших и наибольший относительный объем фальсификаций.

В табл. 10, 11 и 12 представлены официальные итоги голосования по избирательным округам и результаты реконструкции, округленные до целых голосов, а также получаемые на их основе итоги выборов.

В качестве официальных результатов здесь взяты именно цифры с сайта избирательной комиссии, которые несколько отличаются от данных стенограмм. Если мелкие расхождения можно отнести на счет ошибок при подсчете бюллетеней, то крупные, скорее всего, являются результатом целенаправленных действий. Так, для участка 215 суммарное число голосов, поданных за провластных кандидатов, завышено по сравнению со стенограммой на 41, а за прочих кандидатов – занижено на 10 (невозможность вбросить бюллетени еще не спасает от фальсификаций). Аналогичные цифры для участка 220 составляют 57 и 25 соответственно. Кроме того, для участка 216 занижены обе суммы – на 19 и 10 голосов. Эти манипуляции данными почти не влияют на общую картину, но примечательны сами по себе.

Кандидаты в таблицах отсортированы по убыванию их официального результата в округе. Жирным шрифтом выделены 5 наибольших значений в каждой колонке. В подвалах таблиц приведены сводные данные по партийной принадлежности победителей. Хотя для отдельных участков она демонстрирует

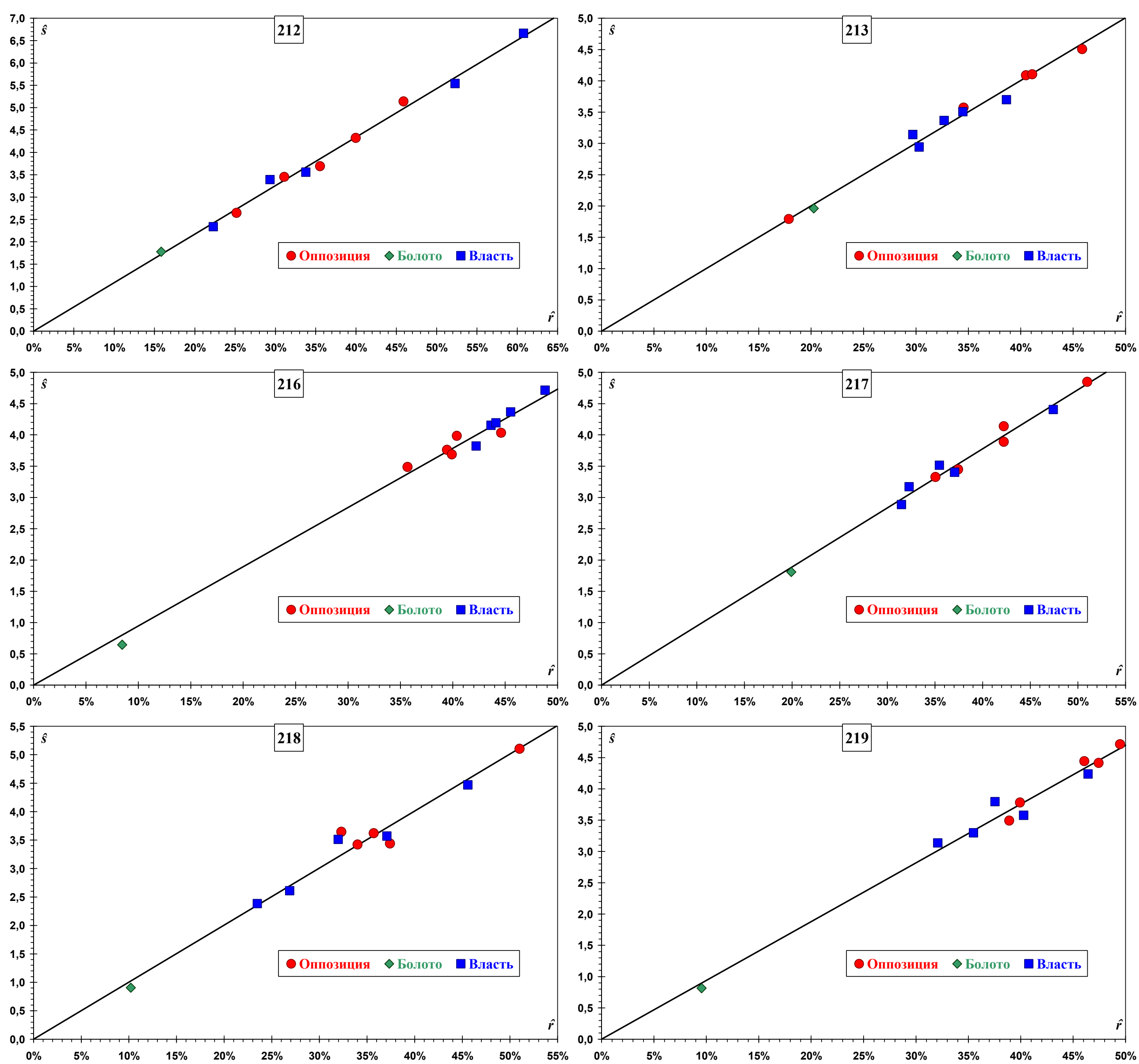

Рис. 14. Сумма условных вероятностей голосования за кандидата в зависимости от его результата – реконструированные данные некоторый разнобой, суммарная картина по округам очень стабильна. В каждом из трех округов по итогам только лишь дистанционного голосования победили по 3 кандидата от ЕР и по 2 от КПРФ. Если его итоги добавить к реконструированными итогами бумажного голосования, то всюду получаются по 2 победителя от ЕР и по 3 от КПРФ. Как уже было отмечено, дистанционно голосует преимущественно административно-зависимый электорат, более лояльный власти. Поэтому общий результат оппозиции не может быть хуже ее результата на ДЭГ. В реальности он оказался даже лучше – 9 мест (из 15) против 6.

Также следует отметить, что если среди официальных победителей, представленных исключительно провластными кандидатами, имеются по 1 кандидату от ЛДПР и СРЗП, локально примкнувших на этих выборах к ЕР, то по результатам реконструкции они не проходят в Совет.

Таблица 9. Параметры реконструкции итогов голосования

| Участок | 212 | 213 | 214 | 215 | 216 | 217 | 218 | 219 | 220 |
|---|---|---|---|---|---|---|---|---|---|
| $n$ | 437 | 256 | 388 | 473 | 318 | 290 | 577 | 355 | 346 |
| $\delta$ | 23,5 | 88,1 | 259,9 | 0 | 105,1 | 164,5 | 518,2 | 62,0 | 81,4 |
| $k$ | 10,9 | 10,0 | 9,8 | 9,6 | 9,5 | 9,4 | 10,0 | 9,4 | 10,2 |
| $1 - kk'$ | 0,08% | 0,08% | 0,56% | 0,20% | 0,09% | 0,08% | 0,31% | 0,13% | 0,20% |
| $(\delta+\delta_0)/n$ | 38% | 34% | 67% | 0% | 33% | 57% | 90% | 17% | 24% |

Таблица 10. Сравнение официальных и реконструированных итогов голосования и выборов для избирательного округа №1

| Кандидат | Партия | ДЭГ | Официально | | | | | Реконструкция | | | | |
|---|---|---|---|---|---|---|---|---|---|---|---|---|
| | | | 212 | 213 | 214 | Сумма | +ДЭГ | 212 | 213 | 214 | Сумма | +ДЭГ |
| Володичева | ЕР | **395** | **251** | **162** | **302** | **715** | **1110** | **228** | **106** | 127 | **461** | **856** |
| Кабанов | ЕР | **302** | **274** | **150** | **304** | **728** | **1030** | **266** | 94 | **134** | **493** | **795** |
| Григорян | ЕР | 277 | **191** | **153** | **286** | **630** | **907** | 128 | 89 | 79 | 296 | 573 |
| Клюкин | ЕР | 246 | **204** | **144** | **297** | **645** | **891** | 148 | 81 | 115 | 344 | 590 |
| Лешко | ЕР | **279** | **169** | **147** | **286** | **602** | **881** | 97 | 83 | 91 | 271 | 550 |
| Яблочкина | КПРФ | **284** | 124 | 87 | 64 | 275 | 559 | **201** | **125** | **194** | **520** | **804** |
| Бганцов | КПРФ | **279** | 108 | 75 | 53 | 236 | 515 | **175** | **111** | **164** | **449** | **728** |
| Слепченко | КПРФ | 209 | 84 | 76 | 48 | 208 | 417 | 136 | **112** | **164** | 412 | **621** |
| Устинов | КПРФ | 201 | 96 | 63 | 55 | 214 | 415 | **155** | **94** | **167** | **416** | 617 |
| Бахтинов | ЛДПР | 148 | 130 | 39 | 26 | 195 | 343 | 69 | 55 | 79 | 203 | 351 |
| Шамова | СРЗП | 115 | 69 | 33 | 32 | 134 | 249 | 110 | 49 | 97 | 256 | 371 |
| **Победители** | **ЕР** | **3** | 5 | 5 | 5 | **5** | **5** | 2 | 1 | 1 | **2** | **2** |
| | **КПРФ** | **2** | 0 | 0 | 0 | **0** | **0** | 3 | 4 | 4 | **3** | **3** |
| | **СРЗП** | **0** | 0 | 0 | 0 | **0** | **0** | 0 | 0 | 0 | **0** | **0** |
| | **ЛДПР** | **0** | 0 | 0 | 0 | **0** | **0** | 0 | 0 | 0 | **0** | **0** |

Таблица 11. Сравнение официальных и реконструированных итогов голосования и выборов для избирательного округа №2

| Кандидат | Партия | ДЭГ | Официально | | | | | Реконструкция | | | | |
|---|---|---|---|---|---|---|---|---|---|---|---|---|
| | | | 215 | 216 | 217 | Сумма | +ДЭГ | 215 | 216 | 217 | Сумма | +ДЭГ |
| Шик | ЕР | **299** | **215** | **201** | **224** | **640** | **939** | **199** | **155** | **138** | **492** | **791** |
| Федорив | ЕР | **234** | 184 | **187** | **211** | **582** | **816** | 164 | **140** | 107 | 412 | **646** |
| Серба | ЕР | **231** | 158 | **197** | **209** | **564** | **795** | 157 | **139** | 103 | 399 | 630 |
| Щербович | ЕР | 188 | 154 | **197** | **205** | **556** | **744** | 150 | 134 | 94 | 378 | 566 |
| Вавилова | ЛДПР | 181 | 117 | **202** | **204** | **523** | **704** | 117 | **145** | 91 | 353 | 534 |
| Дубровский | КПРФ | **266** | **224** | 94 | 64 | 382 | 648 | **228** | **142** | **148** | **518** | **784** |
| Соколов | КПРФ | **226** | **201** | 85 | 53 | 339 | 565 | **202** | 127 | **122** | **451** | **677** |
| Богучарская | КПРФ | 207 | **215** | 79 | 53 | 347 | 554 | **215** | 125 | **122** | **463** | **670** |
| Скопинов | КПРФ | 191 | **186** | 85 | 47 | 318 | 509 | **188** | 128 | **109** | **425** | 616 |
| Соцкова | КПРФ | 186 | 174 | 76 | 44 | 294 | 480 | 175 | 113 | 102 | 390 | 576 |
| Булах | ЕР | 131 | 93 | 15 | 25 | 133 | 264 | 95 | 27 | 58 | 180 | 311 |
| **Победители** | **ЕР** | **3** | 1 | 4 | 4 | **4** | **4** | 1 | 3 | 1 | **1** | **2** |
| | **КПРФ** | **2** | 4 | 0 | 0 | **0** | **0** | 4 | 1 | 4 | **4** | **3** |
| | **СРЗП** | **0** | 0 | 0 | 0 | **0** | **0** | 0 | 0 | 0 | **0** | **0** |
| | **ЛДПР** | **0** | 0 | 1 | 1 | **1** | **1** | 0 | 1 | 0 | **0** | **0** |

Таблица 12. Сравнение официальных и реконструированных итогов голосования и выборов для избирательного округа №3

| Кандидат | Партия | ДЭГ | Официально | | | | | Реконструкция | | | | |
|---|---|---|---|---|---|---|---|---|---|---|---|---|
| | | | 218 | 219 | 220 | Сумма | +ДЭГ | 218 | 219 | 220 | Сумма | +ДЭГ |
| Мазирко | ЕР | **325** | **542** | **172** | **190** | **904** | **1229** | **263** | 133 | **124** | **520** | **845** |
| Булкина | ЕР | **284** | **540** | **198** | **195** | **933** | **1217** | **214** | **165** | **147** | **526** | **810** |
| Зеленченко | ЕР | **303** | **537** | **166** | **178** | **881** | **1184** | 185 | 126 | 116 | 426 | 729 |
| Васильева | ЕР | 224 | **534** | **181** | **186** | **901** | **1125** | 155 | **143** | **124** | 422 | 646 |
| Аникеева | СРЗП | 209 | **532** | **153** | **188** | **873** | **1082** | 135 | 114 | 108 | 357 | 566 |
| Ушаков | КПРФ | **253** | 19 | 146 | 118 | 283 | 536 | 186 | **176** | **154** | **516** | **769** |
| Силиванова | КПРФ | **252** | 20 | 138 | 118 | 276 | 528 | **206** | **168** | **156** | **530** | **782** |
| Федотов | КПРФ | 237 | 20 | 133 | 83 | 236 | 473 | 196 | **164** | **131** | 491 | 728 |
| Афонин | КПРФ | 237 | 30 | 116 | 81 | 227 | 464 | **294** | 142 | 112 | **549** | **786** |
| Разина | КПРФ | 174 | 22 | 113 | 88 | 223 | 397 | **216** | 138 | 118 | 472 | 646 |
| Сборнов | ЛДПР | 149 | 6 | 28 | 30 | 64 | 213 | 59 | 34 | 39 | 132 | 281 |
| **Победители** | **ЕР** | **3** | 4 | 4 | 4 | **4** | **4** | 2 | 2 | 3 | **2** | **2** |
| | **КПРФ** | **2** | 0 | 0 | 0 | **0** | **0** | 3 | 3 | 3 | **3** | **3** |
| | **СРЗП** | **0** | 1 | 1 | 1 | **1** | **1** | 0 | 0 | 0 | **0** | **0** |
| | **ЛДПР** | **0** | 0 | 0 | 0 | **0** | **0** | 0 | 0 | 0 | **0** | **0** |

Проверка реконструкции

Альтернативный метод реконструкции итогов голосования дает рассмотрение зависимости частной явки (за или против какого-то кандидата или их группы) от общей. Используемая при этом модель поведения избирателей постулирует линейный характер такой зависимости для социально-однородного региона. С точки зрения выявления фальсификаций здесь наиболее удобен выбор явки против власти в качестве зависимой переменной, поскольку тогда фальсифицированные участки оказываются по одну сторону относительно прямой, на которую ложатся достоверные участки. [7,8]

К сожалению, к рассматриваемым выборам этот подход напрямую неприменим, т.к. он предполагает определение трех параметров, характеризующих политическую ситуацию в регионе, для чего необходимы хотя бы несколько десятков избирательных участков без фальсификаций. Кроме того, использование ДЭГ в значительной степени обессмысливает само понятие явки. К электронному участку приписываются лишь те избиратели, которые заранее предприняли некоторую активность для этого, тогда как все политически пассивные избиратели (в т.ч. абстиненты) остаются приписанными к номерным участкам. Это приводит к аномально высокой дистанционной явке и занижает обычную.

В свете указанных обстоятельств здесь метод был модифицирован таким образом, чтобы позволить пусть и не получить независимую реконструкцию итогов голосования, но хотя бы проверить уже выполненную. Вместо общей явки и явки против власти здесь рассматриваются общее число отметок, сделанных избирателями в бюллетенях, и число отметок, сделанных за кандидатов, не вошедших по официальным итогам выборов в число победителей в своих избирательных округах.

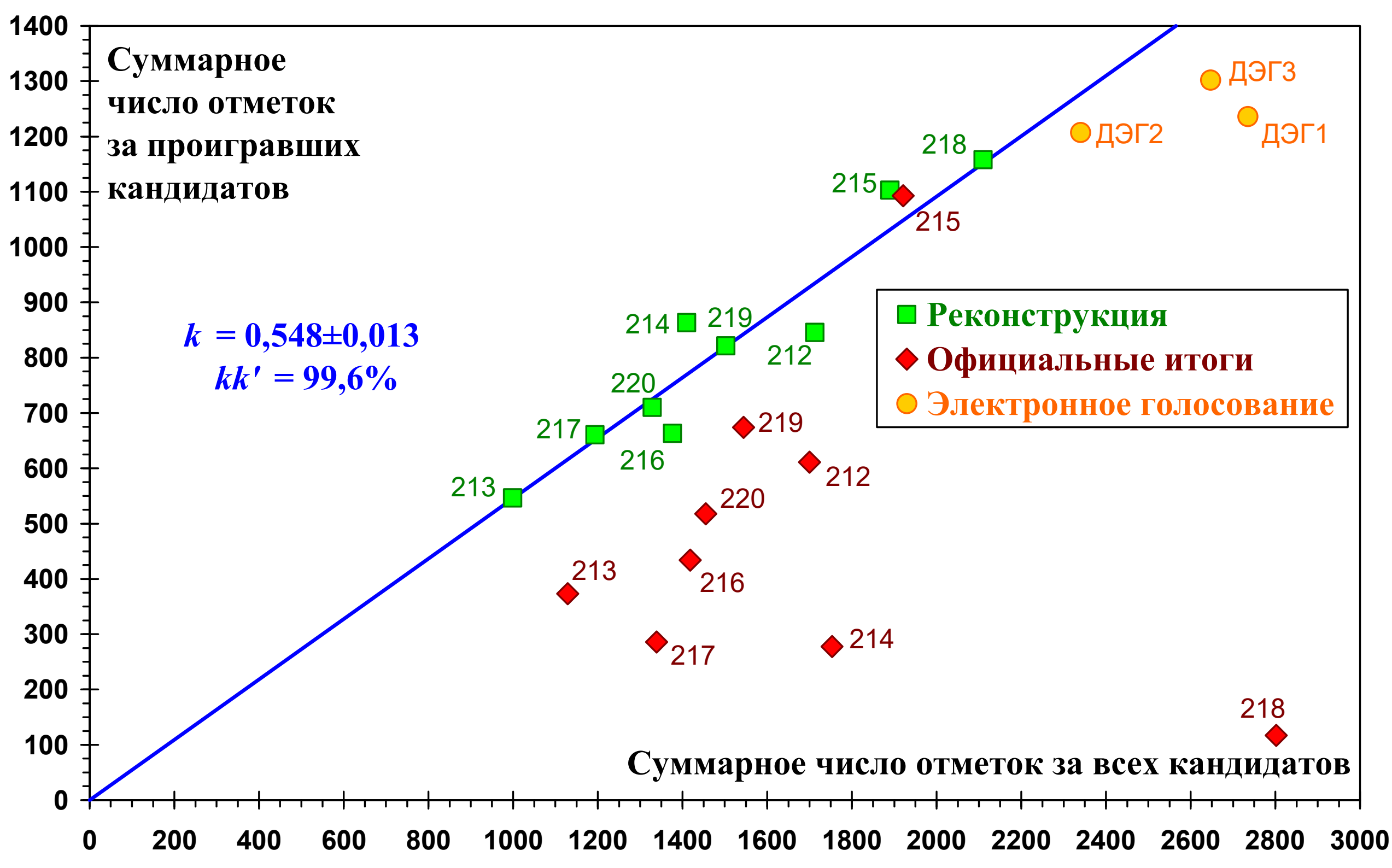

Рис. 15. Проверка реконструкции итогов голосования

Как показывает рис. 15, реконструированные итоги выборов при таком подходе подчиняются сформулированной модели поведения избирателей, тогда как официальные – нет. Реконструированные итоги для обычных участков описываются прямой, проходящей через начало координат, что свидетельствует о незначительной роли административного ресурса на этих выборах [7,8]. При выполнении регрессии точки рассматривались без весов, а как мера качества приближения данных прямой пропорциональностью использовано произведение угловых коэффициентов прямой и обратной зависимостей, которое оказывается весьма близко к единице.

Угловой коэффициент регрессионной прямой лишь немного превышает значение ½, соответствующее равенству голосов за проигравших и победителей. Это говорит об острой конкуренции между основными политическими силами. Поэтому при честном подведении итогов голосования конкретный список победителей должен в значительной степени определяться не столько политической принадлежностью кандидатов, сколько их индивидуальной популярностью. И вполне ожидаемо, что оппозиция может выставить больше активных и харизматичных кандидатов, чем власть.

Официальные итоги для 8 из 9 обычных участков оказываются смещены относительно регрессионной прямой вправо–вниз, как и должно быть при фальсификациях. Вброс голосов за победителей увеличивает абсциссу точки, а переброс голосов от проигравших к победителями уменьшает ее ординату.

Окружные участки ДЭГ тоже немного смещены относительно прямой. Особенно это заметно для округа №1, где удаленно проголосовали более 43% избирателей, принявших участие в выборах, тогда как для округов №2 и №3 – примерно 39%. Административное принуждение, характерное для электронного голосования, вовлекает в него инертных избирателей, которые голосуют за власть либо из послушания, либо из опасения, что их выбор станет известен. Всё это играет роль ослабленных фальсификаций, оказывающих пусть и не качественное, но количественное влияние на итоги голосования.

**Выводы**

Для стенограмм подсчета голосов многомандатных выборов предложены новые формальные методы электоральной статистики. Строгие методы позволяют доказывать наличие фальсификаций, полустрогие – легко его обнаруживать, а нестрогие – реконструировать истинные итоги голосования. Такое богатство результатов обусловлено большим объемом информации, содержащимся в стенограмме подсчета голосов. Это позволяет применять статистические методы для исследования итогов голосования на уровне отдельных участков, что было невозможно при использовании лишь сводных данных.

Последовательности голосов, поданных за или против каждого кандидата, а также последовательности консолидированных бюллетеней в случае вброса (подмены) пачек идентично заполненных бюллетеней содержат аномалии, которые могут быть выявлены инструментами математической статистики. Построенные на их основе тесты количественно определяют степень нашей уверенность в том, что аномалии обусловлены именно фальсификациями.

Использование числовых характеристик, изменяющихся известным образом при усилении фальсификаций, дает более простой подход к определению их масштаба. Однако такие методы являются лишь качественными, а не количественными. Возможность легко сравнить участки друг с другом не избавляет от необходимости дополнительного определения в их ряду реперных точек.

Вариативность участков по типу фальсификаций, их масштабу и степени перемешивания бюллетеней позволяет проверить как устойчивость предложенных методов к попыткам фальсификаторов замаскировать свою деятельность, так и меру понимания ими того, что это следует делать.

Готовность избирателя одновременно поддерживать различных кандидатов позволяет определить их сходство. В рамках модели, предполагающей линейную взаимосвязь результата кандидата с суммой условных вероятностей голосования за него, может быть выполнена реконструкции итогов голосования. Ее результаты успешно проходят независимую проверку сопоставлением общей активности избирателей с их протестной активностью.

Информационная энтропия бюллетеней как мера фальсификаций и алгоритм реконструкции, основанный на сходстве кандидатов, применимы только в случае голосования с множественными отметками. А все прочие предложенные подходы могут использоваться и для анализа стенограмм подведения итогов голосования лишь с одним допустимым вариантом волеизъявления.

## Примечания

Благодарности

Видеозаписи и стенограммы подсчетов голосов были предоставлены кандидатами и наблюдателями от КПРФ, работавшими на избирательных участках.

Доступность исходных данных

Оригиналы видеозаписей и копии использованных документов могут быть предоставлены по обоснованному запросу к В.М. по адресу makarov@vad1.com.

Вклады авторов

А.В.П. выполнил статистическое исследование и написал статью. В.М. координировал сбор и анализ экспериментальных данных.

Конфликты интересов

Авторы заявляют об отсутствии конфликта интересов.

## Литература